\documentclass[12pt]{article}

\usepackage{graphicx}
\usepackage{epstopdf}
\usepackage[centertags]{amsmath}
\usepackage{amssymb}
\usepackage{amsthm}
\usepackage{amsfonts}
\usepackage{ccaption}
\usepackage{cite}
\usepackage{mathrsfs}

\newcommand{\be}{\begin{equation}}
\newcommand{\ee}{\end{equation}}
\newcommand{\beq}{\begin{eqnarray}}
\newcommand{\eeq}{\end{eqnarray}}

\def\sec#1{Section \ref{#1}}
\def\fig#1{Fig.\,\ref{#1}}

\def\eg{{\it e.g.}}

\def\ie{{\it i.e.$\,$}}

\def\CB{{\cal B}}

\def\CH{{\cal H}}

\def\CM{{\cal M}}

\def\T{{\bf T}}
\def\Sp{{\bf S}}

\def\A5S5{{\rm AdS}_5 \times \S^5}

\def\ii{\mathrm{i}}

\newtheorem{theorem}{Theorem}

\theoremstyle{definition}
\newtheorem{remark}{Remark}

\newtheorem*{sthrm}{Staticity theorem}

\title{\bf{On the uniqueness of extremal vacuum black holes}}

\author{ Pau Figueras\footnote{pau.figueras@durham.ac.uk} \ and James Lucietti\footnote{james.lucietti@durham.ac.uk } \\ \\
 \small \sl Centre for Particle Theory, Department of
Mathematical Sciences, \\ \small \sl University of Durham, South
Road, Durham, DH1 3LE, UK
 }

\date{}

\begin{document}

\maketitle

\begin{picture}(0,0)(0,0)
\put(350, 245){DCPT-09/39}
\end{picture}

\vskip1cm
%Abstract
\begin{abstract}
We prove uniqueness theorems for asymptotically flat, stationary,
extremal, vacuum black hole solutions, in four and five dimensions
with one and two commuting rotational Killing fields respectively.
As in the non-extremal case, these problems may be cast as boundary
value problems on the two dimensional orbit space. We show that the
orbit space for solutions with one extremal horizon is homeomorphic
to an infinite strip, where the two boundaries correspond to the
rotational axes, and the two asymptotic regions correspond to
spatial infinity and the near-horizon geometry. In four dimensions
this allows us to establish the uniqueness of extremal Kerr amongst
asymptotically flat, stationary, rotating, vacuum black holes with a
single extremal horizon. In five dimensions we show that there is at
most one asymptotically flat, stationary, extremal vacuum black hole with a connected horizon, two commuting rotational symmetries and given interval structure and angular momenta. We also provide necessary and sufficient conditions  for four and five dimensional asymptotically flat vacuum black holes with the above symmetries to be static (valid for extremal, non-extremal and even non-connected horizons).
\end{abstract}
\thispagestyle{empty} \setcounter{page}{0}
%\end{titlepage}

\newpage
\renewcommand{\thefootnote}{\arabic{footnote}}
%______________________________________

%%%%%%%%%%%%%%%%%%%%%%%%%%%%%%%%%%%%%%%%%%%%

%\tableofcontents

%~~~~~~~~~~~~~~~~~~~~~~~~~~~~~~~~~~~~~~~~~~~~~~~

%__________________________________________________
\section{Introduction}
%__________________________________________________

The black hole uniqueness theorems are one of the most striking
results of four dimensional General Relativity. In the absence of
matter the precise statement may be summarised by: {\it the only
asymptotically flat, stationary, black hole solution of
the Einstein vacuum equations, with a connected non-degenerate
horizon, is the non-extremal Kerr black hole
solution}~\cite{Carter:1971zc, Carter:1973lh,Robinson:1975bv,Mazur:1982db,Mazur:1983dc,Bunting:1983bu, Carter:1985bu}\footnote{We have omitted various technical assumptions, one of the main ones being the assumption of analyticity. See~\cite{Chrusciel:2008js} for a self-contained rigorous
proof which discusses the precise technical conditions and fills in some gaps in the literature.}. This remarkable
uniqueness theorem has been a cornerstone to our understanding of
semi-classical and quantum properties of equilibrium black holes.
This is because it shows that equilibrium black holes are uniquely
specified by their conserved charges (the mass $M$, the angular
momentum $J$), thus providing a simple physical way of identifying
such states within quantum gravity. In fact, so far our most
successful quantum descriptions have been for {\it extremal} black
holes (within string theory). This is because they are simpler objects within quantum
gravity due to the fact they don't radiate. Curiously though, the
classical uniqueness theorems always assumed the black hole is
non-extremal. Thus, it is of interest to fill this gap in the
uniqueness theorems.

The black hole uniqueness theorem has taken many years to
prove
and in fact follows from a number of more specific results each of
which must be proved separately. It is worth reviewing how this
theorem has been proved for non-extremal black holes. Firstly, one
establishes that spatial sections of the horizon must have $\Sp^2$ topology (this
was first argued by Hawking~\cite{Hawking:1971vc, Hawking:1973uf} and later
strengthened using topological censorship~\cite{Chrusciel:1994tr, Galloway:1999bp}). A
stationary black hole, by definition, possesses a Killing field
$\xi$ timelike near infinity. Since $\xi$ must leave the event
horizon invariant it must be tangent to it. This leads to two cases:
either  $\xi$ is normal to the horizon (\ie tangent to the null
geodesic generators) or  $\xi$ is spacelike on the horizon. These
are referred to as non-rotating and rotating respectively.
In the non-rotating case one can prove the solution must be also
static~\cite{Sudarsky:1992ty}. One can then appeal to a theorem that
states that the only static solution in this class is Schwarzschild
\cite{Israel:1967wq, BM} (even without assuming a connected horizon).  The rotating case is more involved. First one
establishes the rigidity theorem which allows one to deduce that a rotating black hole must be axisymmetric~\cite{Hawking:1971vc,Hawking:1973uf,Chrusciel:1996bj,Friedrich:1998wq}. The way this is
proved is to show that the event horizon is a Killing horizon of
some Killing field $\chi$ which is not parallel to $\xi$
(as its rotating). Then one shows that the Killing field $\chi-\xi$
has closed (space-like) orbits, \ie there exists a $U(1)$ isometry.
Having established this, the rotating case reduces to a study of
stationary, axisymmetric vacuum spacetimes, which by definition are
Weyl solutions. Einstein's equations for such solutions may be
recast as an integrable 2d non-linear sigma model with target space
given by the symmetric space $SU(1,1)/S( U(1)\times
U(1))$~\cite{Ernst:1967wx, Mazur:1983vi}. Properties of this sigma
model then allow one to show uniqueness of Kerr within the class of
asymptotically flat, stationary, axisymmetric vacuum solutions. The
main technical tool is the so called Mazur identity (see \eg, \cite{Carter:1985bu}), which together
with a careful study of boundary conditions on the horizon, the axis
of symmetry and asymptotic infinity allows one to establish
uniqueness.

The purpose of this paper is to extend the known uniqueness theorems
to extremal black holes. Even in the 4d case discussed above this
does not seem to have been discussed in the literature. We will
focus on the rotating case. However, we should note that to the
best of our knowledge the non-rotating case has not been fully ruled
out (the existence of static extremal vacuum black holes has been
ruled out~\cite{Chrusciel:2005pa}, however it has not been shown
that a non-rotating extremal black hole must be static). In the
rotating case the rigidity theorem has also been established in the
extremal case \cite{Hollands:2008wn}.
Thus we know that such solutions must also be stationary and axisymmetric.
We will show that the methods used for the uniqueness theorem for non-extremal black holes can be easily adapted to extremal black holes. The
only difference is the 2d space which the sigma model is defined on.
The 2d space in fact can be identified with the orbit space of the
$\mathbb{R} \times U(1)$ isometry group\footnote{Given a manifold $M$ with a group action of a group $G$, the orbit space is the quotient  $M/G$, which is defined by the equivalence relation ``two points in $M$ are equivalent if they are related by some element of $G$".}. This is an orientable simply connected manifold with boundaries and corners. For non-extremal black holes (with one horizon) this
orbit space is homeomorphic to a semi-infinite strip, with the
boundary corresponding to the two rotation axis and the horizon, and
the one asymptotic region corresponding to spatial infinity~\cite{Carter:1973lh}\footnote{It is worth noting that the orbit space in this case is also homeomorphic to the upper half complex plane with the real line corresponding to the boundary and the asymptotic regions spatial infinity.}. For an
extremal black hole (with one horizon) we will show that the orbit space is an infinite strip
with one of the asymptotic regions corresponding to the horizon, the
other corresponding to spatial infinity and the two (disconnected)
boundaries the rotation axis. The asymptotic region corresponding to
the horizon is fully determined by the near-horizon geometry of such
solutions. Therefore the new ingredient required to prove uniqueness
of extremal black holes (for given $M,J$) is a general understanding of their possible near-horizon geometries. This has in fact already been achieved~\cite{Kunduri:2007vf} (furthermore it turns out such vacuum near-horizon geometries are unique~\cite{Haj, LP, Kunduri:2008rs}). This allows us to deduce our
first result: 
\begin{theorem}
 The only four dimensional, asymptotically
flat, stationary and axisymmetric, rotating, black hole solution of
the Einstein vacuum equations, with a connected degenerate horizon
with non-toroidal sections, is the extremal Kerr solution.
\end{theorem}

The assumption of a non-toroidal horizon is used to determine the structure of the orbit space\footnote{It is an assumption of
the uniqueness theorem of the near-horizon
geometry~\cite{Kunduri:2008rs}. In fact it can be shown that
the most general vacuum near-horizon geometry with toroidal horizon
sections is simply the direct product $R^{1,1}\times T^2$~\cite{Kunduri:2008rs}.}, and in
the context of black holes is of course justified by topological
censorship. Similarly, axisymmetry is justified by the rigidity
theorem of~\cite{Hollands:2008wn}. Together with these results, our
Theorem 1 provides a complete classification of rotating
vacuum black holes with a single degenerate horizon.\footnote{We have not spelled out all technical assumptions, the main one being analyticity of the metric. As in the non-extremal case, it would be interesting to remove such assumptions.}

The uniqueness/classification problem for higher dimensional black
holes remains a much tougher challenge. This is shown by the
fact that even in five dimensions explicit examples of
asymptotically flat black holes are known with $\Sp^3$~\cite{MP} and $\Sp^2
\times \Sp^1$~\cite{Emparan:2001wk,Emparan:2001wn,Pomeransky:2006bd} horizon topology, which demonstrates explicitly black hole non-uniqueness.
Again we will focus on asymptotically flat, stationary vacuum black
holes with a single connected component of the horizon. Some general
results are known. Firstly, by generalising Hawking's original
argument, the topology of the horizons sections have been shown to
be of positive Yamabe type~\cite{Galloway:2005mf, Galloway:2006ws}.
Static non-extremal black hole solutions have been shown to be given
by higher dimensional versions of Schwarzschild~\cite{
Gibbons:2002bh}. Furthermore, it has been shown that non-extremal
non-rotating black holes of this kind must be
static~\cite{Rogatko:2005ys}. This leaves the rotating case. To this
end, a generalisation of the rigidity theorem has been established:
stationary, non-extremal, rotating black holes must admit of $U(1)$
isometry with spacelike orbits
\cite{Hollands:2006rj,Moncrief:2008mr}. This guarantees the
existence of two commuting Killing fields. However, the
generalisation of Weyl's class to higher dimensions requires $D-2$
commuting Killing fields~\cite{Emparan:2001wk, Harmark:2004rm}
(something the $D=4$ uniqueness theorem relies on crucially). For
$D>4$ we see this is more than guaranteed by rigidity. Luckily
for $D=5$ asymptotic flatness is consistent with the existence of
$U(1)^2$ isometry with spacelike orbits (this is the maximal abelian
subgroup of the rotation group).\footnote{In fact all known 5d black
hole solutions have $\mathbb R \times U(1)^2$ symmetry.} Therefore
all progress so far has been made by focusing on solutions with this
symmetry as they belong to the Weyl
class\footnote{See~\cite{Chrusciel:2008rh} for some recent rigorous
results.}. In fact an analogue of the 4d uniqueness theorem has
already been proved: {\it asymptotically flat, stationary,
non-extremal black hole vacuum solutions with $U(1)^2$ rotational
symmetry are uniquely specified by the conserved charges $J_1,J_2$
and the interval structure} \cite{Hollands:2007aj}. The interval
structure~\cite{Hollands:2007aj,Hollands:2008fm}, is a set of data
which consists of an ordered set of open intervals which correspond
to boundary segments of the orbit space\footnote{This data is also
known as rod-structure in the earlier
literature~\cite{Emparan:2001wk, Harmark:2004rm}.}. These segments
either correspond to the horizon orbit space, or sets where certain
combination of the rotational Killing fields vanish. Each of these
intervals is specified by its length and, for the parts not
corresponding to the horizon, a 2d vector which determines which
linear combination of rotational Killing fields vanishes on that
interval. Note that a consequence of the theorem is that all other data, such as the mass, is in fact determined for such solutions by the interval structure and angular momentum.

In this paper we will also extend this theorem to extremal black
holes in this class. We will focus on the non-static case (the static
case has been ruled out~\cite{Chrusciel:2005pa}), although we should note that, as in 4d, it has not been shown that non-rotating implies static for this class of solutions. Once again the same method as in
the non-extremal case can be used although the structure of the
orbit space changes. A general understanding of near-horizon geometries of such black holes~\cite{Kunduri:2007vf} allows us to show that, as in 4d, the orbit space is an infinite strip with one
asymptotic region fully determined by the near-horizon geometry. The
interval structure in this case only corresponds to the axes of
rotation, as the horizon orbit space is no longer treated as part of the
boundary of the full orbit space and instead consists of an asymptotic region. Thus our second result is: 
\begin{theorem}
 Consider a five dimensional,
asymptotically flat, stationary black hole solution of the vacuum Einstein equations, with $U(1)^2$ isometry
with spacelike orbits and a connected degenerate horizon (with
non-toroidal sections). There
exists at most one such solution with given angular momenta
$J_1,J_2$ and a given interval structure. 
\end{theorem}

Again, the assumption of a non-toroidal horizon is used to determine the structure of the orbit space -- the general form of a near-horizon geometry~\cite{Kunduri:2007vf} is used for this. In fact near-horizon geometries
for such black holes have been recently
classified~\cite{Kunduri:2008rs}. Naturally, the results are more
complicated than in 4d and one can have three possible near-horizon
geometries: two classes of $\Sp^3$ horizon geometries and one class
of $\Sp^2 \times \Sp^1$ horizon
geometry. Despite this lack of near-horizon uniqueness our Theorem shows that specifying the interval structure and angular momenta is sufficient to uniquely specify a full extremal black hole solution (and therefore also its near-horizon geometry!). 

While the above uniqueness theorems mainly concern stationary but non-static black holes, it is also of interest to make general statements regarding static black holes. In \cite{Carter:1973lh} Carter proved that a stationary and axisymmetric $D=4$ electro-vacuum black hole solution with a non-degenerate connected horizon and zero angular momentum is necessarily static. This theorem does not apply to the case of a horizon with several connected components. In fact for $D=5$ stationary non-static multi black hole states with zero total angular momentum are certainly possible: the Black Saturn \cite{Elvang:2007rd} provides the first explicit example of such a spacetime. Therefore it is of interest to generalise Carter's staticity theorem. We will derive necessary and sufficient staticity conditions for asymptotically flat $D=4,5$ stationary vacuum black hole solutions with $U(1)^{D-3}$ rotational symmetry and a possibly non-connected horizon (whose components may be either degenerate or non-degenerate). Indeed we find that in the case of non-connected horizons, the vanishing of the total angular momentum of the spacetime is not sufficient to guarantee staticity in accordance with the known examples. However, the vanishing of the Komar angular momentum or angular velocity for each component of the horizon is sufficient.

This paper is organised as follows. First, we introduce some preliminary concepts required for the proof and we use them to derive the necessary and sufficient staticity conditions. Then we analyse the orbit space for an extremal black hole, highlighting the differences
to the non-extremal case. Then we prove the uniqueness theorem,
which requires a detailed analysis of the behaviour of a certain
function at asymptotic infinity, the axes and the near-horizon
region. We end with a discussion of our results and provide two
Appendices with the proofs of the staticity theorems and  the Einstein equations written in a certain coordinate system.

\bigskip

\noindent \textbf{Note added:} While this paper was nearing
completion, a paper~\cite{Amsel:2009et} appeared which also proves the
uniqueness of the four dimensional extremal Kerr black hole. After the first version of our paper appeared we were informed of an existing uniqueness proof for extremal Kerr~\cite{Meinel} (that also assumes axisymmetry) which uses a different method.

%__________________________________________________
\section{Stationary vacuum black holes with rotational symmetries}
%__________________________________________________

%__________________________________________________
\subsection{Preliminaries}
%__________________________________________________
\label{sec:Preliminaries}

Consider stationary asymptotically flat black hole spacetimes
$(\CM,g_{\mu\nu})$ which are solutions to Einstein's vacuum
equations $R_{\mu\nu}=0$. We will restrict to $D=4,5$ and assume
there is an  $\mathbb R \times U(1)^{D-3}$ isometry group. For
four-dimensional black holes, the rigidity theorem
\cite{Hawking:1971vc,Hawking:1973uf}, which guarantees the existence
of a $U(1)$ isometry, has recently been extended to extremal
black holes~\cite{Hollands:2008wn}. In five dimensions rigidity only
guarantees the existence of a $U(1)$ isometry and has only been
proved for non-extremal black holes (see~\cite{Hollands:2008wn} for
partial results in the extremal case), and thus the assumption of a
$U(1)^2$ isometry appears to be a genuine restriction. We assume
that spatial sections of the event horizon are connected $(D-2)$-dimensional
compact (closed and orientable) manifolds which we denote by $\CH$.
Therefore $\CH$ inherits an $U(1)^{D-3}$ isometry group which
defines an (effective) $U(1)^{D-3}$ action on $\CH$ which constrains
its topology as follows. In $D=4$ we must have $\Sp^2,\T^2$, whereas
in $D=5$ we must have $\Sp^3,\Sp^2\times \Sp^1,
L(p,q),\T^3$ (see \eg, \cite{Gowdy, Chrusciel2rot}).\footnote{So far, all known asymptotically flat black hole solutions with
topology $L(p,q)$ are singular \cite{lenses}. }  In four dimensions
only $\Sp^2$ is allowed by topological censorship
\cite{Galloway:1999bp}. In five dimensions only $\Sp^3$ (and
$L(p,q)$) and $\Sp^2 \times \Sp^1$ horizon sections are
allowed~\cite{Galloway:2005mf, Galloway:2006ws}. Thus in all cases
we will assume non-toroidal horizon sections. It is worth noting
that the orbit space $\CH/U(1)^{D-3}$ is then always a compact
interval with a certain combination of the Killing fields vanishing
at the endpoints.

We will denote the stationary Killing field by $\xi$ and the
rotational Killing fields by $m_i$ where $i=1, \dots, D-3$. Thus we
have a vacuum solution with $D-2$ commuting Killing fields $(\xi,
m_i)$ and hence belongs to the (generalised) Weyl class
\cite{Emparan:2001wk}. As is well known for such solutions one can
always find coordinates $(\xi^\alpha, x^a)$ such that the spacetime
metric is \be ds^2= \label{weylform} G_{\alpha\beta}(x) d\xi^\alpha
d\xi^\beta+ g_{ab}(x)dx^a dx^b \ee where $\alpha, \beta$ run over
$(0,i)$ and $\xi=
\partial /\partial \xi^0$ and $m_i= \partial /\partial \xi^i$ and $a,b$ run over the 2d base
space $\mathcal{B}$. The vacuum Einstein equations are then equivalent to the following equations on $\mathcal{B}$:
\begin{eqnarray}
&&D_a( \rho\, G^{-1} D^a G)=0  \label{Gequation} \\
&&R_{ab}= D_aD_b \log \rho-\frac{1}{4}\textrm{Tr} \, ( D_a G^{-1} D_b G) \label{Riccibase}
\end{eqnarray}
where $\rho^2 \equiv -\det G \geq 0$, and  $D_a$ and $R_{ab}$ are the metric connection and Ricci tensor associated to the base metric $g_{ab}$. Note that $G$ is a matrix of scalar fields on $\mathcal{B}$ with components $G_{\alpha \beta}$ and we will suppress these matrix indices. We will find it useful to write the Killing part of the metric $G_{\alpha\beta}$ as
\be
\label{GmetricD}
G_{\alpha\beta} d\xi^\alpha d\xi^\beta= -\frac{\rho^2}{\det\lambda}\,dt^2
    +\lambda_{ij}(d\phi^i-\omega^i\,dt)(d\phi^j-\omega^j\,dt)
\ee
where we have introduced coordinates adapted to the Killing fields $\xi=\partial/\partial t$ and $m_i=\partial /\partial \phi^i$.

Define the twist one-forms
\begin{equation}
 \Omega_i=\star\left(m_1\wedge\dots\wedge m_{D-3}\, \wedge dm_i\right)\,,
 \label{eqn:Omegai}
\end{equation}
where  $\star$ is the Hodge dual in the full spacetime (we choose
the orientation to be
$\epsilon_{t\phi_1\dots\phi_{D-3}ab}=+\rho\,\epsilon_{ab}^{(2)}$.)
The Einstein vacuum equations imply that the $\Omega_i$ are closed,
so locally we can find the corresponding twist potentials $Y_i$ so
that $\Omega_i=dY_i$. In fact, Einstein's equations also imply that
$\xi \cdot \Omega_i$ is a constant function and thus assuming at
least one $m_i$ vanishes somewhere in space-time (as is the case for
asymptotically flat spacetimes) it follows that $\xi \cdot
\Omega_i=0$. Therefore we see the $\Omega_i$ are also closed
one-forms on spatial sections $\Sigma_t$ of the spacetime. In fact
topological censorship guarantees that for
asymptotically flat black holes the exterior of the black hole
(domain of outer communications) is simply connected -- it follows
that the functions $Y_i$ exist globally (\ie $\Omega_i$ are exact).
Note that $Y_i$ are only defined up to an additive constant: this is
the only gauge freedom associated to them. Now we can state an
important fact: at a fixed point of any of the rotational Killing
fields one must have $\Omega_i=0$ for {\it all} $i$, and thus $Y_i$
are constant on these subspaces.

Using the explicit parametrisation
introduced above one can check that
\begin{equation}
dY_i=\frac{\det\lambda}{\rho}\lambda_{ij}\star_{2}d\omega^j\,, \label{twistD}
\end{equation}
which can be inverted:
\begin{equation}
 d\omega^i=-\frac{\rho}{\det\lambda}\,\lambda^{ij}\star_{2}dY_j\,,\label{domegaD}
\end{equation}
where $\lambda^{ik}\lambda_{kj}=\delta^{i}_{\phantom i j}$.  In
terms of the twist potentials equation \eqref{Gequation} is
equivalent to\footnote{From \eqref{GmetricD} it appears that
\eqref{Gequation} yields four sets of equations for $\lambda_{ij}$
and $Y_i$. However, one can show that only two of them are
independent and these are the ones that we display.}
\begin{subequations}
\label{lambdaYeqs}
\begin{align}
&D^a(\rho\, \lambda^{ik} \, D_a \lambda_{kj})=-\frac{\rho}{\det\lambda}\,\lambda^{ik}\, (D^a Y_k)(D_a Y_j)\\
&D^a \left(\frac{\rho}{\det \lambda} \, \lambda^{ij} \, D_a Y_j
\right)=0 \; .
\end{align}
\end{subequations}
It is worth noting that for $D=4$, in which case there is only one
rotational Killing field $m_1$, we have two scalar functions
$X \equiv \lambda_{11}$ and $Y\equiv Y_1$. The above equations can then be
written as a single equation for the complex Ernst potential
$E=X+\ii\, Y$~\cite{Ernst:1967wx}.

We will now show that the event horizon must be a Killing
horizon (see \eg,\cite{Weinstein1}).\footnote{Since we are assuming the existence of rotational
symmetries this is easy to prove and is sometimes referred to as the
``weak rigidity theorem''. For $D>4$, the strong rigidity theorem
(\ie without assuming the existence of rotational symmetries) has
only been proved for non-extremal black holes.}  Consider the
hypersurface $\mathcal{N}=\{ \rho=0,\, \det\lambda>0 \}$. Since
$\det G=-\rho^2$ we immediately see that the induced metric on
$\mathcal{N}$ is degenerate with signature $(0,+,\ldots,+,+)$.
Therefore $\mathcal{N}$ is a null hypersurface. Consider the vector
field $\xi+\omega^i\, m_i$ which is tangent and null on
$\mathcal{N}$. It follows it is also normal to $\mathcal{N}$. From
(\ref{domegaD}) we see that the $\omega^i$'s are  constant on
$\mathcal{N}$ and therefore the normal $\xi+\omega^i\, m_i$ is also
Killing on $\mathcal{N}$. Hence we deduce that $\mathcal{N}$ is a
Killing horizon of the Killing vector field $\chi \equiv
\xi+\omega^i|_{\mathcal{N}}\, m_i$. The trace of the extrinsic
curvature associated to the normal $\chi$ must vanish since it is a
Killing vector, and thus $\mathcal{N}$ is the apparent horizon. From
standard results (see \eg, \cite{Hawking:1973uf}) for asymptotically
flat, stationary spacetimes the apparent horizon coincides with the
event horizon and therefore we deduce that $\mathcal{N}$ is the
event horizon. We will assume the spacetime does contain a black
hole, \ie\ the set $\mathcal{N}$ is not empty. Note that the angular velocities of the horizon are defined by the constants $\omega_H^i \equiv \omega^i|_{\mathcal{N}}$ and the black hole is referred to as non-rotating if $\omega_H^i=0$.

So far we have
treated extremal and non-extremal solutions on equal footing. We can
now we distinguish them as follows. One can compute
\begin{equation}
\chi^2= -\frac{\rho^2}{\det\lambda} +\lambda_{ij}\,(\omega^i- \omega^i|_{\mathcal{N}})\,(\omega^j-\omega^j|_{\mathcal{N}})\,,
\end{equation}
from which it follows that
\begin{equation}
d\chi^2\big|_{\mathcal{N}}= -\frac{d\rho^2}{\det\lambda}\bigg|_{\mathcal{N}}
\end{equation}
The surface gravity $\kappa$ is defined by $d\chi^2|_{\mathcal{N}}=-2\,\kappa\, \chi|_{\mathcal{N}}$ and thus we see that the horizon is extremal if and only if $d\rho^2|_{\mathcal{N}}=0$. Unfortunately this equation does not seem to be a useful way of imposing extremality.\footnote{ Although note that this implies that $\rho^2$ is not a good coordinate on a degenerate horizon, whereas it is in  the non-degenerate case \cite{Carter:1973lh}.}

Similarly we may also define axes of rotation (\ie, the set of fixed
points of the rotational Killing fields) as the set $\mathcal{A}=\{
\rho=0, \det\lambda=0, 0<\rho^2/\det\lambda<\infty \}$. Since
$\lambda_{ij}=m_i \cdot m_j$ we see that on $\mathcal{A}$ the $m_i$ are
not all linearly independent and hence the twist 1-forms
$\Omega_i=dY_i=0$. Thus, since $Y_i$ are continuous (in fact smooth) functions
on the whole spacetime, they are constants on each connected component of a $\mathcal{A}$. For a black hole spacetime with a connected horizon the set $\mathcal{A}$ has two disconnected components $\mathcal{A}^{\pm}$, so $\mathcal{A}=\mathcal{A}^+ \cup \mathcal{A}^-$ with $\mathcal{A}^{\pm}$ connected, each of which is connected to asymptotic infinity (see next section)\footnote{Note this does not assume that the same linear combination of $m_i$ vanishes everywhere on each of $\mathcal{A}^{\pm}$.} . Therefore it follows that the value of the constants $Y_i^{\pm} \equiv Y_i|_{\mathcal{A}^{\pm}}$, which thus determine $Y_i$ on the whole of $\mathcal{A}$, can be fixed by comparing to their values in the asymptotically flat region of the spacetime (it turns out they are related to the angular momenta of the spacetime). This fact is important in the uniqueness proofs.

Before moving on to the uniqueness theorems, we will derive simple necessary and sufficient conditions for such black holes to be static, \ie\  for the stationary Killing field $\xi$ to be hypersurface orthogonal. First note that staticity is equivalent to the functions $\omega^i \equiv 0$. Therefore for this class of black hole solutions it is clear that static implies non-rotating (\ie\ $\omega_H^i=0$). In fact the converse statement  (\ie\ non-rotating implies static) is also true in this case and can be deduced from the staticity theorem we prove below. Furthermore, Carter proved that for $D=4$ non-extremal black holes in the above class, zero angular momentum implies the spacetime is static~\cite{Carter:1973lh}: our staticity theorem also  generalises this result. We are now ready to state our result (see Appendix \ref{appendix:static} for proof):  
\begin{sthrm}
 Consider an asymptotically flat, stationary, $D=4,5$ black hole solution of the vacuum Einstein equations, with a $U(1)^{D-3}$ isometry with spacelike orbits, a connected non-degenerate or degenerate horizon with compact sections of non-toroidal topology. Such spacetimes are static if and only if $\omega^i_H J_i=0$ (where $\omega^i_H$ are the angular velocities of the horizon and $J_i$ the angular momenta). 
\end{sthrm}

\begin{remark}
We deduce that sufficient conditions for staticity are either $\omega^i_H=0$ (non-rotating horizon) or $J_i=0$ (zero angular momentum). Note that in 4d one of these conditions is also necessary. This allows a complete classification of the class of black holes specified in the above theorem. In the non-extremal case it has been shown that staticity implies the solution is Schwarzschild~\cite{Israel:1967wq, BM, Gibbons:2002bh}. In the extremal case, static vacuum near-horizon geometries must be direct products of $\mathbb R^{1,1}$ and a Ricci flat $D-2$ compact manifold~\cite{Chrusciel:2005pa} -- this is not compatible with the horizon topologies and symmetries we are considering and hence there can be no extremal static black holes of this kind. 
\end{remark}

\begin{remark}
\label{remark:multistatic}
It is interesting to note how the above theorem changes if one drops the assumption of a connected horizon. In particular vanishing of the total angular momentum of the spacetime is no longer sufficient. Indeed multi black hole spacetimes in this class, which are non-static but have zero angular momentum, are known explicitly (\eg, the Black Saturn~\cite{Elvang:2007rd}). The reason the above result fails in this case is that the axes set $\mathcal{A}$ has more than two disconnected components one of which is necessarily \textit{not} connected to asymptotic infinity -- this means one cannot fix the $Y_i$ everywhere on $\mathcal{A}$ in terms of the angular momenta. In fact, if in the above staticity theorem we drop the assumption of a connected horizon and instead assume we have a horizon with $I=1,\dots N$ components (which may be either degenerate or non-degenerate in any combination), it is possible to prove that the spacetime is static if and only if
\begin{equation}
   \sum_{I=1}^N \omega_I^i J^I_i =0\,, \label{eq:multiplestat}
\end{equation}
where $\omega_I^i$ are the angular velocities and $J^I_i$ are the Komar angular momenta of the $I^{\textrm{th}}$ horizon. The proof is given in Appendix \ref{appendix:static}. Notice that the condition \eqref{eq:multiplestat} may be satisfied if $J_i^I=0$ or $\omega^i_I=0$ for all $I=1, \dots N$. However, it follows that  the vanishing of  the {\it total} angular momenta of the spacetime $\sum_{I=1}^N J^I_i$ is not sufficient to guarantee staticity of the spacetime. \end{remark}

%__________________________________________________
\subsection{Orbit space}
%__________________________________________________
\label{sec:orbit}

Due to the symmetries of the spacetime, the natural space to work on
is the orbit space $\CM/[\mathbb R \times U(1)^{D-3}]$ (\ie the
space of orbits of the isometry group, also known as the factor
space).\footnote{More precisely one considers the orbit space of the domain of outer communication of the black hole. In general one expects this to be an orbifold,
however due to topological censorship it is in fact a
manifold~\cite{Hollands:2008fm}.} In the present context, the
results of the previous section show that this can be identified
with the 2d base space $\mathcal{B}$ which is defined by the
integrable subspace orthogonal to the Killing fields. Thus $\mathcal{B}
\cong \CM/[\mathbb R \times U(1)^{D-3}]$ and we deduce that the
orbit space has a metric $g_{ab}$. So far we have not introduced
explicit coordinates on $\mathcal{B}$. It is well known that  $\rho$
is harmonic on $\mathcal{B}$, \ie $D^aD_a \rho=0$. This allows one
to show that in the region $\rho>0$ one can use $\rho$ as a
coordinate on $\mathcal{B}$ (\ie it has no critical points
\cite{Carter:1973lh,Chrusciel:2008js}). Furthermore one can use the
harmonic conjugate function $z$, defined by $dz=-\star_2d\rho$, as
the other coordinate so \be g_{ab}(x)dx^a dx^b =
e^{2\nu(\rho,z)}(d\rho^2 +dz^2) \ee for some function $\nu(\rho,z)$.
These coordinates are always valid outside the horizon and away from the axes, $\rho >0$,
irrespective of whether the horizon is extremal or not. It is
important to understand the global structure of the orbit space.

\subsubsection{Non-extremal case}
\begin{figure}[t]
\begin{center}
\includegraphics[scale=0.9]{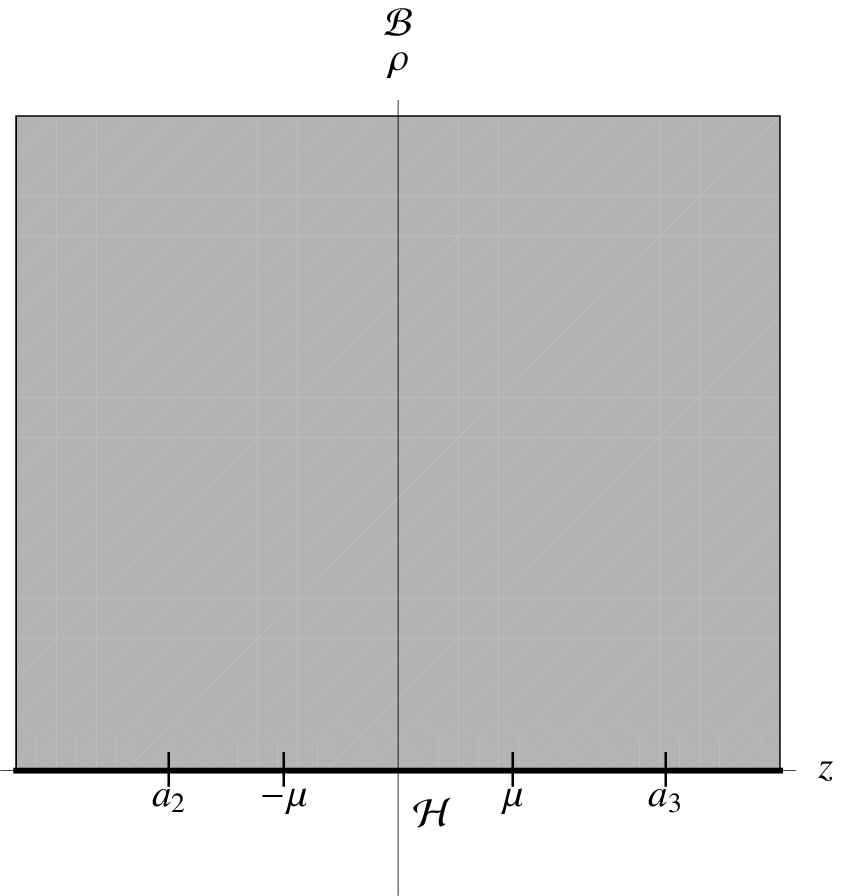}
\hspace{0.5cm}
\includegraphics[scale=0.9]{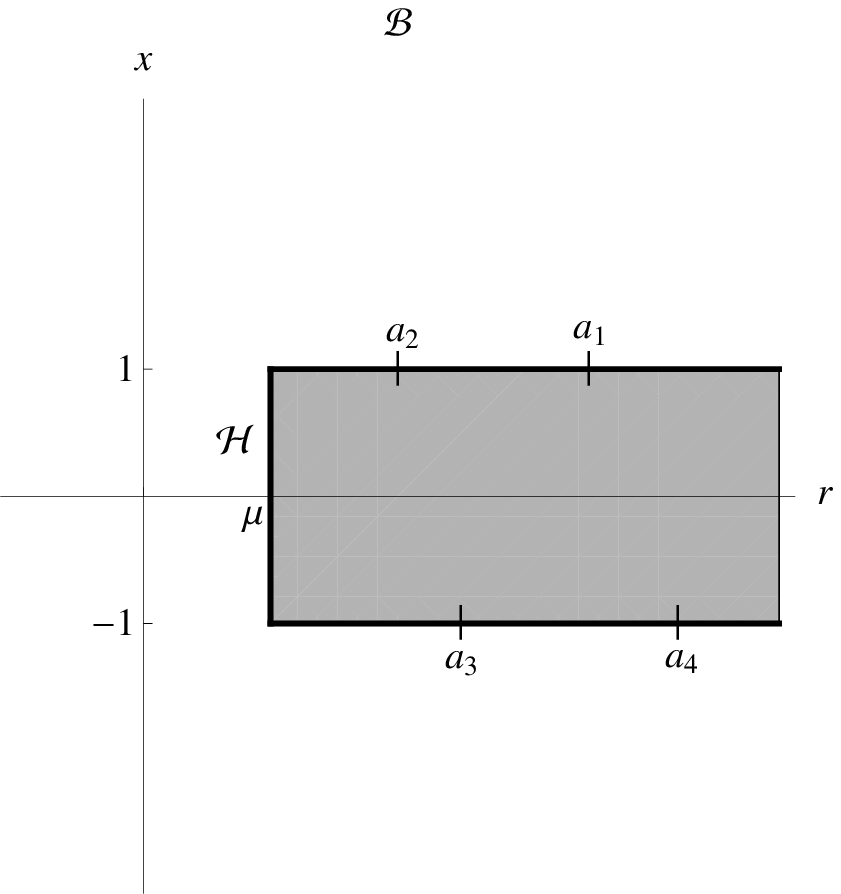}
\caption{Orbit space $\mathcal{B}$ for non-extremal black holes. \textit{Left:} $\mathcal{B}$ can be mapped to the upper half of the complex plane $\zeta=z+\ii\,\rho$. The boundaries and corners correspond to intervals on the real axis. \textit{Right:} $\mathcal{B}$ can also be represented as a semi-infinite strip in the $w=r+\ii\,x$ plane.}
\label{fig:NonExt}
\end{center}
\end{figure}

In the non-extremal case it has been shown that
$\mathcal{B}$ is a 2d orientable simply connected manifold with a connected boundary
with corners~\cite{Carter:1973lh, Hollands:2007aj,Hollands:2008fm} . The boundary consists of a union of segments which
correspond to the fixed point sets of the rotational Killing fields or the horizon orbit space, joined by the corners. In 4d there is one rotational Killing field and thus the boundary segments are specified simply by their length. In 5d the boundary segments which do not correspond to the horizon are specified by a length and a 2d vector which determines which linear combination of the rotational Killing fields $m_i$ vanishes -- this is referred to as the {\it interval structure} (also known as rod data)~\cite{Hollands:2007aj,Hollands:2008fm} (see also
\cite{Harmark:2004rm}). The 2d vectors of neighbouring boundary segments are also required to satisfy a compatibility condition. It has been shown that given the interval structure the spacetime manifold $\CM$ may be reconstructed from the orbit space up to diffeomorphism. Furthermore, $(\rho,z)$
are global coordinates on the orbit space\footnote{Although note that $\rho$ is \textit{not} a good coordinate on the horizon of the full spacetime metric, whereas in fact $\rho^2$ is \cite{Carter:1973lh}.   } showing that $\mathcal{B}$ is conformal to
the upper half of the complex plane $\zeta=z+\ii\,\rho$, with the
boundary and corners mapped on to a set of intervals on the real
axis $\rho=0$. Assuming that spatial sections of the horizon $\mathcal{H}$ are connected and non-toroidal (which we will from now
on), the part of the boundary representing the horizon is simply an
open interval on the $z$-axis, say $(-\mu,\mu)$.  One can in fact
show that the length of this interval is $2\mu = \kappa
A_H/(2\pi)^{D-3}$ where $A_H$ and $\kappa$ are the area and surface gravity of the horizon~\cite{Carter:1973lh,Hollands:2007aj}. As is well known~\cite{Carter:1973lh} one can conformally map this to
the semi-infinite strip $r \geq \mu$ and $|x| \leq 1$ in the
$w=r+\ii\,x$ plane using $\rho^2=(r^2-\mu^2)(1-x^2)$ and $z=r\,x$,
with $r=\mu$ and $x=\pm 1$ corresponding to the horizon and axes
respectively. More precisely the subsets of the orbit space corresponding to $\mathcal{H}$ and $\mathcal{A}^{\pm}$ are $ \{ r=\mu, \; |x| \leq 1 \}$ and $\{ r>\mu,\; x=\pm 1 \}$ respectively. In these coordinates the base metric reads
\be
\label{semistrip}
g_{ab}dx^adx^b=e^{2\nu}(r^2-\mu^2 x^2) \left( \frac{dr^2}{r^2-\mu^2}+ \frac{dx^2}{1-x^2} \right) \; .
\ee
The structure of $\CB$ in both the complex
$\zeta$-plane and in the complex $w$-plane is depicted in
\fig{fig:NonExt}.

% Figure

\subsubsection{Extremal case}
In the extremal case the structure of $\mathcal{B}$ changes.  Any
point outside the horizon is at an infinite proper distance away
from the horizon. Therefore the horizon actually corresponds to
another asymptotic region rather than a part of the boundary. We will show that the orbit space of the exterior of an extremal black hole with a connected horizon is homeomorphic to an infinite strip, such that the two disconnected boundaries correspond to the axes and the two asymptotic regions correspond to spatial infinity and the horizon.  As in the non-extremal case the
boundaries are split into segments characterised by their interval
structure (their length and in 5d a 2d vector which determines which
combination of rotational Killing fields vanish).  In terms of the standard
$(\rho,z)$ coordinates, we will argue below that the horizon is a
point and in fact $\CB$ is conformal to the upper half of the
$\zeta$-plane with the point on the real axis corresponding to the
horizon removed (which has the same topology as an infinite strip).

To determine the topology of the orbit space for a vacuum extremal
black hole we can exploit some general results which have been
derived for their corresponding near-horizon geometries.
In~\cite{Kunduri:2007vf} it was shown that the near-horizon limit of
any stationary extremal black hole with $U(1)^{D-3}$ rotational
symmetries (and non-toroidal horizon sections), in a large class of theories
which include vacuum gravity, can be written as \be
ds^2=\Gamma(\sigma) [ -C^2r^2dv^2+2dvdr]+ \frac{d\sigma^2}{\det
\gamma(\sigma)} +
\gamma_{ij}(\sigma)(d\phi^i+k^irdv)(d\phi^j+k^jrdv)\,, \ee
 where $\Gamma$ is a strictly positive function and $C,k^i$ are constants. The constant $C$ is actually a trivial parameter (due to a scaling symmetry) and may be set to any convenient value -- this is a freedom we will exploit shortly. Note
that the horizon is at $r=0$ and spatial sections are $(D-2)$-dimensional compact manifolds with $(\sigma, \phi^i)$ coordinates.
The Killing field $\chi=\partial /\partial v$ is tangent to the null
geodesic generators of the horizon. The Killing fields $m_i=\partial
/\partial \phi^i$ generate the $U(1)^{D-3}$ isometry and $\sigma$
takes values in a closed interval corresponding to the orbit space of $\mathcal{H}$ (see~\cite{Kunduri:2008rs}). Since the near-horizon geometry is a vacuum solution with $D-2$
commuting Killing fields $(\chi, m_i)$ one can always find
coordinates $(\xi^\alpha, x^a)$ to write the metric in the Weyl form (\ref{weylform}) with  $\chi=
\partial /\partial \xi^0$ and $m_i= \partial /\partial \xi^i$. Explicitly, for $r>0$ define coordinates \be t=v+\frac{1}{C^2 r}, \qquad
\bar{\phi}^i=\phi^i+k^iC^{-2}\log r \ee in which the near-horizon geometry
becomes \be ds^2= \Gamma(\sigma)\left[ -C^2r^2dt^2 +\frac{dr^2}{C^2r^2}
\right] +\frac{d\sigma^2}{\det \gamma(\sigma)} +
\gamma_{ij}(\sigma)(d\bar{\phi}^i+k^irdt)(d\bar{\phi}^j+k^jrdt) \ee
which allows us to read off the base metric \be g_{ab}(x)dx^a dx^b
=\Gamma \left( \frac{dr^2}{C^2r^2}+ \frac{d\sigma^2}{Q} \right) \ee
where $Q \equiv \Gamma \gamma$, and the Killing part of the metric
\be G_{\alpha \beta} d\xi^\alpha d\xi^\beta = -\Gamma(\sigma) C^2r^2
dt^2+\gamma_{ij}(\sigma)(d\bar{\phi}^i+k^irdt)(d\bar{\phi}^j+k^jrdt)
\; . \ee
Recall that the function $\rho$ is defined by $\rho^2 \equiv
-\det G$. In this case we therefore have $\rho^2= C^2 r^2
\Gamma \gamma= C^2 r^2Q$. The vacuum equations imply that $\rho$ is
harmonic on the base space and this allows one to introduce the
harmonic conjugate function $z$ via $dz=-\star_2 d\rho$. One can check that $\star_2
d\rho = -rC^2 d\sigma + \frac{1}{2} \dot{Q}dr$ where we use an
orientation defined by $\epsilon_{r\sigma}>0$ and ``dots" refer to
$\sigma$ derivatives. It follows that $\rho$ is harmonic (\ie
$\,d\star_2 \rho=0$) if and only if $\ddot{Q}+2C^2=0$. Note that this
agrees exactly with what was found in the classification of vacuum
near-horizon geometries in~\cite{Kunduri:2008rs}. Then, the equation for
$z$ can be integrated to give $z=-\frac{1}{2} r\dot{Q}$, where we have
set the integration constant to zero. Since $Q \geq 0$ with equality only at two isolated points which define the boundary of the horizon orbit space~\cite{Kunduri:2008rs}, we integrate to get
$Q=C^2(-\sigma^2+\sigma_0^2)$ where $\sigma_0$ is a positive constant so $-\sigma_0 \leq \sigma \leq \sigma_0$ (wlog the linear term has been set to zero by translating the $\sigma$
coordinate).  Now define a coordinate $0\leq \theta \leq \pi$ by $\cos\theta
=x=\sigma/\sigma_0$ so $Q=C^2\sigma_0^2(1-x^2)= C^2\sigma_0^2
\sin^2\theta$ and the 2d base metric is simply \be g_{ab}(x)dx^a dx^b =
\frac{\Gamma}{C^2r^2} \left( dr^2+r^2 d\theta^2 \right)\,, \ee with
$\rho=C^2\sigma_0\, r\sin\theta$ and $z=C^2\sigma_0\, r \cos\theta$. We will now exploit the scaling freedom associated to the constant $C$ to set $C^2\sigma_0=1$. Thus
$d\rho^2+dz^2 = dr^2 +r^2 d\theta^2$ and hence \be
e^{2\nu}= \frac{\Gamma}{C^2r^2} \; .\ee

Thus we have showed that for any vacuum near-horizon geometry the
horizon $r=0$ is located at the origin of the upper half plane
$\rho=z=0$, and the coordinate $r$ (which originates from the
Gaussian null coordinates) is simply the polar coordinate
$r=\sqrt{\rho^2+z^2}$. Furthermore, the coordinate $x \equiv \cos\theta$, so $|x|\leq 1$,
which parametrises the horizon orbit space corresponds to the polar
angle $\tan\theta=\rho/z$ of the $(\rho,z)$ plane. However, this
description is potentially misleading as the proper length of the
horizon segment is in fact finite and the proper distance from any
point to the horizon is infinite (since the conformal factor is
singular at the origin of the upper half plane). A better
description of the base space in this region is achieved by defining
a coordinate $y=\log r$ so $-\infty <y<\infty$ and \be
g_{ab}(x)dx^adx^b = \frac{\Gamma(x)}{C^2} \left( dy^2+
\frac{dx^2}{1-x^2} \right) \; . \ee This shows that the base space
in the near-horizon limit is conformal to the strip $|y|<\infty$ and
$|x|\leq 1$. This space has two disconnected boundaries given by $x
=\pm 1$, which correspond to the axes, and two asymptotic ends,
namely the horizon, which is at $y \to -\infty$, and spatial
infinity, which is at $y \to \infty$. Note that in these coordinates it is easy to see that the proper length of
the horizon orbit space is $l_H= \int_{0}^\pi d\theta \sqrt{\Gamma}
C^{-1}>0$. In terms of the original Weyl coordinates $(\rho,z)$, the
orbit space is homeomorphic to the upper half of the complex plane
with a point which corresponds to the location of the horizon (say
the origin) removed.

\begin{figure}[t]
\begin{center}
\includegraphics[scale=0.9]{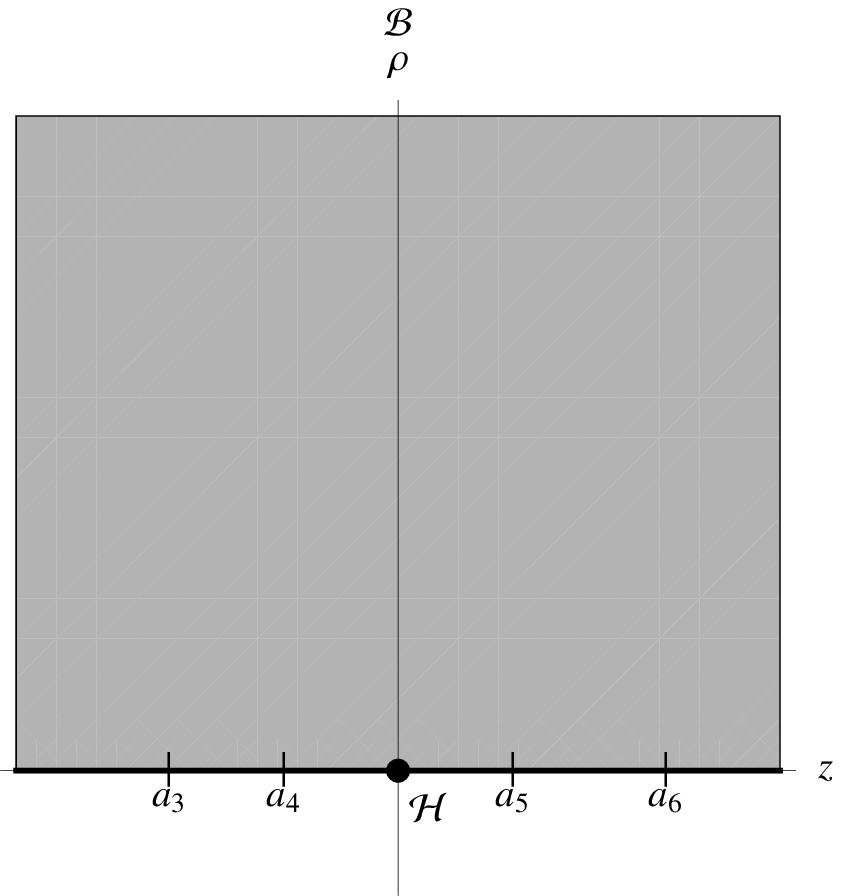}
\hspace{0.5cm}
\includegraphics[scale=0.9]{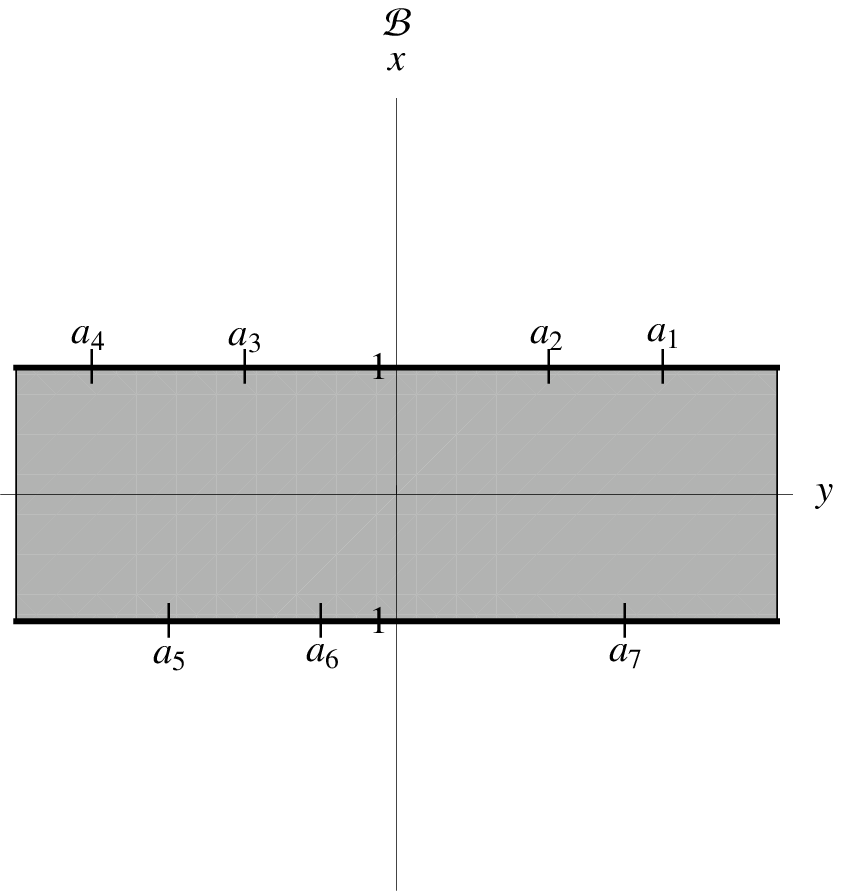}
\caption{Orbit space $\mathcal{B}$ for extremal black holes. \textit{Left:} $\mathcal{B}$ can be regarded as the upper half of the complex plane $\zeta=z+\ii\,\rho$, with the point on the boundary corresponding to the the horizon (the origin) removed. \textit{Right:}  $\mathcal{B}$ is better viewed as an infinite strip in the $y+\ii\,x$ plane with $r=\log y$. The horizon (not shown) is at $y\to -\infty$ and spatial infinity is at $y\to+\infty$. The two disconnected boundaries $(x=\pm 1)$ correspond to the axes.}
\label{fig:Ext}
\end{center}
\end{figure}

We may use this result to deduce the structure of the orbit space
for the full extremal black hole (\ie not just its near-horizon
geometry). Cutting the spacetime off outside the horizon, we see
that the region connected to asymptotic infinity is identical to
that for a non-extremal black hole. The region connected to the
horizon is well approximated by the near-horizon geometry provided
we cut sufficiently near the horizon. Thus gluing together these
shows that the orbit space for the full extremal black hole must
also be homeomorphic to an infinite strip with the horizon and
asymptotic infinity located at the two asymptotic ends and the axes
are the two boundaries. The structure of the orbit space for
extremal black holes is depicted in \fig{fig:Ext}. We note that for
$\rho>0$ the functions $(r,\theta)$ defined by
$\rho=r\sin\theta$ and $z=r \cos\theta$, are good coordinates
(since ($\rho,z$) are). As shown above $r \to 0$ corresponds to the
asymptotic region near the horizon ($r=0$ is not part of the orbit
space) as it must coincide with the near-horizon limit. For the full black hole $r \to \infty$ corresponds to asymptotic infinity.
Therefore $(r,\theta)$ are global coordinates on the orbit space of the exterior of the full extremal black hole whose metric is then\footnote{Notice that if one takes the zero horizon interval length limit of the non-extremal case $\mu \to 0$, the coordinates in the semi-infinite strip description of the orbit space of a non-extremal black hole (\ref{semistrip}) coincide exactly with the polar coordinates of the $\rho,z$ plane.} \be g_{ab}dx^a
dx^b= e^{2\nu(r,\theta)} (dr^2 +r^2d\theta^2) \ee with $r>0$ and
$0\leq \theta \leq \pi$, and the boundary consists of $\{ r>0, \; \theta=0 \}$
and $\{ r>0, \; \theta =\pi \}$ which correspond to the two disconnected parts of the axes $\mathcal{A}^{\pm}$.

% Figure
%

%__________________________________________________
\section{Uniqueness proof}
%__________________________________________________

Both the classic uniqueness theorem in $D=4$, as well as its more recent extension to $D=5$, rely on a number of remarkable properties of the Einstein vacuum equations for Weyl solutions. In particular, one can recast the equation of motion \eqref{lambdaYeqs} as
\begin{equation}
 D_a\left(\rho\,\Phi^{-1}D^a\Phi\right)=0 \label{eqsigmaD}
\end{equation}
where
\begin{equation}
 \Phi=
\left(
\begin{array}{cc}\displaystyle
 \frac{1}{\det\lambda} & \displaystyle-\frac{1}{\det\lambda}\,Y_i\\ & \\
\displaystyle-\frac{1}{\det\lambda}\,Y_j &
\displaystyle \lambda_{ij}+\frac{1}{\det\lambda}\,Y_i\,Y_j
\end{array}
\right)\,,
\end{equation}
is a real, $(D-2)\times (D-2)$, {\it positive-definite}, symmetric matrix with unit determinant. It is useful to define the current $J= \Phi^{-1} d\Phi$ which satisfies
the conservation law $D^a( \rho \, J_a)=0$, and the ``square-root'' matrix $S$ such that $\Phi= S^T S$ and $\det S=1$.

Now, consider two asymptotically flat vacuum black hole solutions
$(\CM_{[0]},g_{[0]})$ and $(\CM_{[1]},g_{[1]})$ with a connected
degenerate horizon and denote all data with the corresponding
subscripts. Assume the two solutions have identical interval
structure and angular momenta. We showed that the
orbit space for any such black hole is homeomorphic to an infinite
strip or equivalently the upper half $z+i\rho$ plane minus the
origin. Therefore we may identify the orbit spaces and hence can
assume $\rho_{[0]}=\rho_{[1]}$ and $z_{[0]}=z_{[1]}$. As in the
non-extremal case, the orbit space together with the interval
structure allows one to reconstruct the spacetime manifold together
with the $\mathbb{R} \times U(1)^{D-3}$-action up to diffeomorphism
and thus we may set $\CM_{[0]}=\CM_{[1]}$ and $\xi_{[0]}=\xi_{[1]}$,
$m_{[0]\, i}=m_{[1]\, i}$.  It now remains to show uniqueness of the
spacetime metric.

First, consider the two corresponding solutions of (\ref{eqsigmaD}), $\Phi_{[0]}$ and $\Phi_{[1]}$ and define the deviation matrix $\Psi
\equiv\Phi_{[1]}\Phi_{[0]}^{-1}-\mathbf{1}$, where $\mathbf{1}$
denotes the identity matrix. Using (\ref{eqsigmaD}) one can verify the Mazur identity
\begin{equation}
 D_a(\rho\,D^a\textrm{Tr}\,\Psi)={\rho}\,
\textrm{Tr}\left( N^T_a N^a\right)\,
\label{mazur}
\end{equation}
where $N \equiv S_{[1]}(J_{[0]} -J_{[1]})S_{[0]}^{-1}$.
Explicitly
\begin{equation}
\textrm{Tr}\,\Psi =-1+\frac{\det\lambda_{[0]}}{\det\lambda_{[1]}}
    +\lambda_{[0]}^{ij}\left(\lambda_{[1]ij}-\lambda_{[0]ij}\right)
    +\frac{1}{\det\lambda_{[1]}}\,\lambda_{[0]}^{ij}
    \left(Y_{[1]i}-Y_{[0]i}\right)\left(Y_{[1]j}-Y_{[0]j}\right)\,.
\label{tracePsi}
\end{equation}
The Mazur identity is the key ingredient to proving uniqueness. We will follow the argument of~\cite{Weinstein2, Hollands:2008fm} closely.
It can be shown that $\textrm{Tr}\,\Psi \geq 0$~\cite{Hollands:2008fm}.\footnote{This is easy to see in $D=4$ as
$\textrm{Tr}\, \Psi= \frac{1}{X_{[0]}X_{[1]}}[
(X_{[1]}-X_{[0]})^2+(Y_{[1]}-Y_{[0]})^2]$.} If we now define an auxiliary $\mathbb{R}^3-\{ 0 \}$ by $(x,y,z)= (\rho \cos \phi, \rho \sin\phi, z)$ where $(\rho,z)$ are the coordinates on the orbit space (recall $\rho=z=0$ is not part of this space for an extremal black hole) the Mazur identity can be written as
\begin{equation}
\nabla^2\, \textrm{Tr}\,{\Psi} \geq 0
\end{equation}
on $\mathbb{R}^3 - \{ z\textrm{-axis} \}$ where $\nabla$ is the
vector derivative and $\textrm{Tr}\, \Psi$ is an axisymmetric
function on $\mathbb{R}^3 -\{ 0 \}$.

The uniqueness proof works as follows. Suppose $\textrm{Tr}\,\Psi$
is bounded on $\mathbb{R}^3 - \{ 0 \}$ and vanishes somewhere. Then
by the maximum principle\footnote{In fact for the domain $\mathbb{R}^3 - \{ z\textrm{-axis} \}$, one needs a modification of the standard maximum principle on $\mathbb{R}^3$. The necessary result is proved in~\cite{Chrusciel:2007ak} Proposition C.4. We thank Piotr Chrusciel for pointing this out.}  $\textrm{Tr}\, \Psi \equiv 0$ everywhere in
$\mathbb{R}^3 - \{ 0 \}$ and hence in $\mathcal{B}$.
Since $\textrm{Tr}\, \Psi= \textrm{Tr}\, (F^TF) -(D-2)$, where
$F=S_{[1]}S_{[0]}^{-1}$ and $\det F=1$, it follows that $F\equiv
\mathbf{1}$ and hence $\Phi_{[0]} \equiv\Phi_{[1]}$, \ie we must
have $\lambda_{[0] ij}= \lambda_{[1] ij}$ and $Y_{[0]i }=Y_{[1]i}$
everywhere which establishes uniqueness of the solution
$(\lambda_{ij},Y_i)$. Now, work in $(r,\theta)$ coordinates on
$\CB$. Since the
conformal factor $e^{2\nu}$ cancels from equation (\ref{domegaD}), it determines $\omega^i$ (up to a
constant which may be fixed from the fact that $\omega^i \to 0$ at spatial infinity) . Finally, the
metric on the base space $\CB$ is then also determined by
(\ref{Riccibase}) as it reduces to a first order PDE for $\nu$ (see
Appendix \ref{subsec:sphpolarcoords}) which is integrable and has a unique solution (up to some
constant which may be fixed from the spatial asymptotics). Thus the whole
problem of uniqueness boils down to showing that
$\textrm{Tr} \, \Psi$ is bounded everywhere on the orbit space and vanishes somewhere. We will do this next by working in $(r,x)$ coordinates, where $x=\cos\theta$, throughout.\footnote{We should note that the $(r,x)$ coordinates are also valid for non-extremal black holes in the region $r>0$ and thus our analysis of asymptotic infinity and the axes is also valid in this case.}

%__________________________________________________
\subsection{Asymptotic infinity}
%__________________________________________________

Let us consider asymptotic spatial infinity. In both 4 and 5
dimensions we demand strict asymptotic flatness. Asymptotic spatial
infinity in polar coordinates then corresponds to $r\to \infty$. In 4d asymptotic
flatness implies \be \lambda_{11}= r^2(1-x^2)[ 1+ O(r^{-1})], \qquad
Y_1= 2\,J\,x(3-x^2)+ O(r^{-1}) \ee as $r \to \infty$, where $J$ is
the ADM angular momentum. Note that we have chosen a ``symmetric''
gauge for $Y$, \ie $Y_{x=1}=-Y_{x=-1}$. It is then easy to see that
two solutions with the same $J$ must satisfy $\textrm{Tr}\, \Psi=
O(r^{-2})$ as $r\to \infty$.

In 5d asymptotic flatness implies that as $r\to \infty$
\begin{align}
&\lambda_{11} = r(1+x)\left[1+O(r^{-1})\right], \quad \lambda_{22}= r(1-x)\left[1+O(r^{-1})\right]\,,\nonumber\\
&\lambda_{12}= (1-x^2)\,O(r^{-1}) \,,\nonumber \\
&Y_1 = y_1- \frac{J_1\, x(2-x)}{\pi} +O(r^{-1}) \qquad Y_2=y_2 - \frac{J_2\, x(2+x)}{\pi} +O(r^{-1})
\label{asym5d}
\end{align}
where $J_i$ are the ADM angular momenta and $y_i$ are integration
constants (\ie we will not fix the gauge for $Y_i$ yet). Using
\eqref{asym5d} in \eqref{tracePsi}, it follows that for two
solutions with the same $J_i$ (and in the same gauge $y_i$), we must
have $\textrm{Tr}\, \Psi =O(r^{-1})$ as $r \to \infty$.

Thus both in 4d and 5d we see that $\textrm{Tr}\, \Psi \to 0$ as
$r\to \infty$ thus proving that not only it is bounded near infinity
but also vanishes there. Recall the fact that $\textrm{Tr} \, \Psi$
vanishes somewhere is a necessary part of the uniqueness theorem.

%__________________________________________________
\subsection{Axes}\label{sec:axes}
%__________________________________________________

We now turn to the boundary conditions near the axes of rotation $\mathcal{A}= \mathcal{A}^+ \cup \mathcal{A}^-$, which in polar coordinates are given by $\mathcal{A}^{\pm}= \{ r>0, \; x=\pm 1 \}$.  These boundary conditions are chosen to guarantee regularity of the spacetime metric near the fixed points of the rotational Killing fields $m_i$.

In 4d there is only one rotational Killing field $m_1$ and regularity requires
\be
\lambda_{11}= a(r)(1-x^2) +O(1-x^2)^2, \qquad G_{t1}= O(1-x^2)
\ee
where $a(r)>0$ and the axis is approached as $x \to \pm 1$. These equations may be expressed in the more elegant equivalent form
\be
\label{4daxesreg}
(1-x^2) \partial_x \log \lambda_{11} = \mp 2 +O(1-x^2), \qquad Y_1= \pm 4J +O(1-x^2)^2
\ee
as $x \to \pm 1$. Note that since all components of the axes of rotation set $\mathcal{A}$ are connected to infinity (recall this follows from our assumption of a connected horizon) we can fix $Y_1$ everywhere on $\mathcal{A}$ by using the asymptotics for $Y_1$ near infinity. This shows that for two solutions with the same $J$ we have $\textrm{Tr} \, \Psi = O(1)$ near the axis.

In 5d we have two Killing fields $m_i$ which can vanish, although by regularity they can only do so simultaneously at isolated points (which correspond to the corners in the orbit space). Consider the part of the axes set $\mathcal{A}^-$ defined by $x=-1$ and suppose we are near infinity where (wlog) $m_1=0$. Regularity then requires that as $x \to -1$
\be
\lambda_{11}=a(r)(1+x)+O(1+x)^2 , \qquad \lambda_{12}=O(1+x), \qquad \lambda_{22}= b(r)+O(1+x)
\ee
where $a(r),b(r)>0$ and $G_{t1}=O(1+x)$ and $G_{t2}=O(1)$ (the latter two equations are equivalent to $\omega^i=O(1)$). Analogous expressions are valid for the behaviour near other parts of $\mathcal{A}^-$ (where potentionally a different combination of the Killing fields vanishes) and also on $\mathcal{A}^+$ where we take $m_2=0$ near infinity. The conditions on $\lambda_{ij}$ may be more elegantly written in the form
\be
\label{5daxesreg}
(1\pm x)^2\textrm{Tr}\,[ (\lambda^{-1}\partial_x \lambda)^2]=1+O(1\pm x), \qquad (1\pm x)\partial_x \log \det \lambda =\mp 1+O(1\pm x)
\ee
as $x \to \mp 1$. In this form one does not need to specify which combination of Killing fields vanishes and thus these conditions provide a more convenient statement of regularity near all parts of $\mathcal{A}^{\pm}$ (not just near infinity). The conditions on $G_{ti}$ can be translated to conditions on $Y_i$. These can be written in a symmetric form by fixing the constant term in $Y_i$ to match with the asymptotics at infinity and picking the gauge $y_1=-J_1/\pi$ and $y_2=J_2/\pi$:
\be
Y_i=\mp \frac{2J_i}{\pi} + O(1 \pm x) \label{Yiaxes}
\ee
as $x \to \mp 1$.\footnote{Note that if we are on a part of an axis where some combination of $m_i$ vanishes, then the corresponding combination of $Y_i$ has an error term which vanishes as $O((1\pm x)^2)$.} As argued earlier, the twist potentials $Y_i$ are constant everywhere on the axes $x=\pm 1$, and therefore equation \eqref{Yiaxes} fixes these constants not only on the parts of the axes that are near asymptotic infinity but everywhere on $\mathcal{A}^{\pm}$ and thus $\mathcal{A}$ (recall since we are assuming a connected horizon all connected components of $\mathcal{A}$ are connected to infinity). Using these conditions one may check that for two solutions with the same $J_i$, near the axes set $\mathcal{A}$ we have  $\textrm{Tr}\, \Psi =O(1)$.

Thus, we have shown that both in 4d and 5d, regularity of the
spacetime metric implies that $\textrm{Tr}\,\Psi=O(1)$ near $x =\pm
1$ thus showing this function is bounded near the parts of the
$z$-axis away from the origin as required.

%__________________________________________________
\subsection{Near the horizon}
%__________________________________________________

It remains to consider the behaviour of $\Psi$ near the horizon.
This is the only part of the uniqueness proof where one really needs to
distinguish between extremal and non-extremal black holes. We will
assume the existence of a single degenerate horizon. As argued earlier, in this case one can use the polar coordinates $(r,\theta)$ defined by $\rho=r\sin\theta$ and $z=r \cos\theta$ with the horizon at $r \to 0$. In the Appendix \ref{subsec:sphpolarcoords} we give expressions for all the
equations of motion in the $(r,\theta)$ coordinates and again for convenience we will use $x=\cos\theta$.

Regularity of the horizon requires that the matrix components $\lambda_{ij}$ and the potentials $Y_i$ have the following expansion near $r=0$:
\begin{equation}
 \lambda_{ij}(r,x)=\lambda_{ij}(x)+O(r)\,,\qquad Y_i(r,x)=Y_i(x)+O(r)
\end{equation}
for some smooth functions $\lambda_{ij}(x)$ and $Y_i(x)$\footnote{Since $r=0$ is not strictly part of the orbit space we do not need to demand regularity at the origin of the auxiliary $\mathbb{R}^3$. This is why the first order terms are $O(r)$ and not $O(r^2)$.}. In fact this is sufficient to establish that $\textrm{Tr} \Psi=O(1)$ as $r \to 0$ and $|x| \leq 1$ (\ie even on the axes). This is easy to see in 4d. Regularity near the axis means $\lambda_{11}(r,x)=(1-x^2)f(r,x)$ and $Y_{[0]}-Y_{[1]}=(1-x^2)g(r,x)$ for smooth functions $f,g$ such that $f>0$ everywhere. It then follows that $\textrm{Tr} \Psi= f_{[0]}^{-1}f_{[1]}^{-1}[ (f_{[1]}-f_{[0]})^2+(g_{[1]}-g_{[0]})^2]$ which is indeed bounded as $r \to 0$ for any $x$. A similar argument can be made in 5d although since the formulas are more cumbersome we do not give details. This completes the proof of Theorem 1 and Theorem 2. $\Box$

For completeness we will now show how the Einstein equation near $r=0$ imply that the functions $\lambda_{ij}(x),Y_i(x)$ obey a set of coupled ODEs which are equivalent to the classification problem of all possible near-horizon geometries in this class -- this has already been solved~\cite{Kunduri:2008rs}. 
At the lowest order, the Einstein equations are (see Appendix \ref{subsec:sphpolarcoords})
\begin{subequations}
\label{EeqsxD}
\begin{align}
&\left[(1-x^2)\lambda^{ik} \lambda_{kj}'\right]'=-\frac{\lambda^{ik}}{\det \lambda } \,Y_k'\,Y_j' \, , \\
&\frac{d}{dx} \left[ \frac{(1-x^2) \lambda^{ij} Y_j'}{\det \lambda} \right]=0\,,\\
 &\partial_r\nu=-\frac{1}{8\,r}\,(1-x^2)^2\left[
\left(\ln\det\lambda\right)'^2+\textrm{Tr}\left((\lambda^{-1}\lambda')^2\right)
+\frac{2}{\det\lambda}\,\lambda^{ij}\,Y_i'\,Y_j'
\right] +O(1)\,,\\
&\partial_x\nu=-\frac{1}{2}\left(\ln\det\lambda\right)'-\frac{1}{8}\,x(1-x^2)\left[
\left(\ln\det\lambda\right)'^2+\textrm{Tr}\left((\lambda^{-1}\lambda')^2\right)
+\frac{2}{\det\lambda}\,\lambda^{ij}\,Y_i'\,Y_j' \right] +O(r)\,,
\end{align}
\end{subequations}
where the prime $'$ denotes the derivative with
respect to $x$.

The second equation in \eqref{EeqsxD} gives us a first integral for
the system of equations for $(\lambda_{ij},Y_i)$
\begin{equation}
\frac{(1-x^2) \lambda^{ij} Y_j'}{\det \lambda} =-k^i \label{thekis}
\end{equation}
where $k^i$ are constants. The significance of these constants is
revealed by looking at the equations for $\omega^i$ which as $r \to
0$ are
\begin{equation}
\partial_r \omega^i=\frac{(1-x^2) \lambda^{ij} Y_j'}{\det \lambda}
+O(r), \qquad \partial_x \omega^i =O(r^2) \; .
\end{equation}
Solving these gives
\begin{equation}
\omega^i=-k^i r +O(r^2) \;
\end{equation}
and we have set the integration constant to zero which corresponds to working in co-rotating coordinates.

We now turn to solving the equations for $\nu$. First note that
using the first two equations in \eqref{EeqsxD}, a straightforward
calculation shows that
\begin{equation}
 (1-x^2)^2\left[
\left(\ln\det\lambda\right)'^2+\textrm{Tr}\left((\lambda^{-1}\lambda')^2\right)
+\frac{2}{\det\lambda}\,\lambda^{ij}\,Y_i'\,Y_j'
\right]=2\,\beta\,,
\label{eqbetaD}
\end{equation}
where $\beta$ is a positive constant. We can now integrate the last two equations in
\eqref{EeqsxD} to obtain
\begin{equation}
 e^{2\nu}=\frac{K}{\det\lambda}\, \left(\frac{1-x^2}{r^2} \right)^{\beta/4} \left[1+O(r) \right]\,,
\end{equation}
for some (positive) constant $K$. We now show that the constant
$\beta$ is fixed by regularity. Using \eqref{thekis}, we can rewrite
\eqref{eqbetaD} as
\begin{equation}
 (1-x^2)^2\left[(\ln\det\lambda)'^2+\textrm{Tr}\left((\lambda^{-1}\lambda')^2\right)
\right]+2\,\det\lambda\,\lambda_{ij}\,k^i\,k^j=2\,\beta\,.
\end{equation}
We can evaluate this expression on any axis, $x=\pm 1$. From the
boundary conditions (\ref{5daxesreg}) as $x \to \pm 1$, the third term on the LHS of this expression
vanishes and the contributions from the two other
terms gives $\beta=4$.

Now, collecting these results, we have shown that as $r \to 0$ the
space-time metric is
\begin{eqnarray}
ds^2 &=& \frac{(1-x^2)}{\det \lambda} \left[ -r^2dt^2 +
\frac{(K+O(r)) dr^2}{r^2} \right] + \frac{(K +O(r))dx^2}{\det
\lambda} \nonumber \\
&+&\lambda_{ij}[ d\phi^i+(k^ir+O(r^2)) dt][ d\phi^j+(k^jr+O(r^2))
dt] \; .
\end{eqnarray}
To examine regularity of the metric at $r=0$ change
coordinates to $v=t-\sqrt{K}/r$ and
$\varphi^i=\phi^i-k^i\sqrt{K}\log r$ which gives
\begin{eqnarray}
ds^2 &=& \frac{(1-x^2)}{\det \lambda} \left[ -r^2dv^2 +
2\sqrt{K}dvdr +O(r^{-1})dr^2 \right] + \frac{(K +O(r))dx^2}{\det
\lambda} \nonumber \\
&+&\lambda_{ij}[ d\varphi^i+(k^ir+O(r^2)) dv][
d\varphi^j+(k^jr+O(r^2)) dv] \; .
\end{eqnarray}
Note that the only singular term is $g_{rr}=O(r^{-1})$; by shifting $v \to v +O(\log r)$ one should be able to prove that $r=0$ is a regular degenerate Killing
horizon of the Killing field $\partial/\partial v$. This requires a higher order calculation which we will not pursue and we will simply assume that we have a regular horizon. In any case, one can
define the near-horizon limit $r \to \epsilon\, r$ and $v \to
v/\epsilon$ taking $\epsilon \to 0$. This gives
\begin{equation}
 ds^2 =
\Gamma(x)[-r^2dv^2+2\sqrt{K}dvdr] + \frac{K \Gamma}{(1-x^2)} dx^2+
\lambda_{ij}(x)(d\varphi^i+k^irdv)(d\varphi^i+k^irdv)
\end{equation}
where we have defined $\Gamma=(1-x^2)/\det \lambda$. Rescaling $v
\to v/\sqrt{K}$, redefining $k^i \to \sqrt{K} k^i$ and setting
$K=C^{-2}$ puts it in exactly the form as the near-horizon
geometries of~\cite{Kunduri:2007vf} (discussed in
\S\ref{sec:orbit}).\footnote{Recall such near-horizon geometries necessarily have an $SO(2,1)$ isometry group.} In order to determine the near-horizon geometry
fully it remains to solve for $\lambda_{ij}(x)$. Our equation for
$\lambda_{ij}$ now reads
\begin{equation}
[(1-x^2)\lambda^{ik}\lambda_{kj}']'=  - \frac{k^ik_j}{\Gamma}
\end{equation}
where we have defined $k_i \equiv \lambda_{ij}k^j$. It is easy to
check this equation is identical to the $ij$ component of the
near-horizon equation given in~\cite{Kunduri:2008rs} (written in the
$\sigma$ coordinate of~\cite{Kunduri:2008rs}). The solutions to this
equation where completely determined in~\cite{Kunduri:2008rs} under
the assumption that spatial sections of the horizon are compact and
non-toroidal. We may therefore simply appeal to the results
of~\cite{Kunduri:2008rs}.

In four dimensions it was shown~\cite{Haj, LP, Kunduri:2008rs} that
there is a unique near-horizon geometry which corresponds to:
\begin{equation} \label{X0Y0}
\begin{aligned}
  \lambda_{11}&=\frac{4 a^2(1-x^2)}{1+x^2}\,,\\
Y_1&=\frac{8a^2\,x}{1+x^2}\,,
\end{aligned}
\end{equation}
where $a>0$ is some integration constant (and we have chosen the symmetric gauge for $Y$ again). This is the near-horizon
geometry of the extremal Kerr solution \cite{Zaslavsky:1997uu, Bardeen:1999px}. We may
fix the constant $a$ as follows. Near the axis the behaviour of $Y_1$ is given by equation (\ref{4daxesreg}), which upon comparison to (\ref{X0Y0}) implies $a^2=J$. Therefore it follows that two asymptotically flat
extremal black hole solutions with the same $J$ must have isometric
near-horizon geometries (this of course also follows from our Theorem 1!). This implies that we must have $\lambda_{[0]11}(x)-\lambda_{[1]11}(x)=O(r)$ and $Y_{[0]1}(x)-Y_{[1]1}(x)=O(r)$, and therefore $\textrm{Tr}\Psi=O(r)$ (for all $|x| \leq 1$). Note that this is in fact stronger than what we needed to establish Theorem 1; in other words knowledge of the near-horizon uniqueness theorem is sufficient but apparently not necessary to prove Theorem 1.

In five dimensions the situation is more complicated. As shown
in~\cite{Kunduri:2008rs} there are three different types of
near-horizon geometry. There is a two parameter family with $\Sp^3$
horizon section topology which is in fact isometric to the
near-horizon geometry of the extremal Myers-Perry black hole. By
comparing $Y_i$ near the axes these two parameters maybe be related
to $J_1,J_2$ thus fully fixing the near-horizon geometry in terms of
the conserved charges. The other two types of near-horizon geometry
are both three parameter families, one with $\Sp^3$ horizon topology
and one with $\Sp^2 \times \Sp^1$ topology. Again, comparing the
$Y_i$ near the axis allows one to fix two combinations of these
parameters and thus in both cases there is one extra parameter. Despite this lack of near-horizon uniqueness, in view of our Theorem 2 we know that given an interval structure and angular momenta one can have at most one asymptotically flat extremal black hole. As a result its near-horizon geometry is also determined uniquely and therefore this extra parameter must be determined by this data.

%__________________________________________________
\section{Discussion}
%__________________________________________________
\label{sec:discussion}

In this paper we have proved uniqueness theorems for asymptotically
flat, stationary, rotating, extremal, vacuum black hole solutions in $D=4,5$
dimensions with $D-3$ commuting rotational symmetries and a single
connected component of the horizon. We have employed the same
methods used in the non-extremal case, both for the classic 4d
uniqueness theorem and the more recent 5d
theorems~\cite{Hollands:2007aj,Hollands:2008wn}. As in the non-extremal case, we have shown that such extremal solutions are uniquely specified by the angular momenta and the interval structure. The interval
structure consists of a set of ordered intervals which correspond to
the boundary segments of the orbit space where the rotational
Killing fields have fixed points. These intervals possess a length
and also a $D-3$ vector that specifies what linear combination
of the rotational Killing fields vanishes there. In $D=4$ the interval structure (for a single black hole) is trivial\footnote{In the non-extremal case there is also an interval corresponding to the horizon and its length (which vanishes in the extremal limit) can be related to the mass and angular momentum.}. Hence in 4d we have shown that such black hole solutions are uniquely specified by their angular momentum and therefore are exhausted by the extremal Kerr family. This fills in an important gap
in the classic 4d uniqueness theorems. 

A new feature arising in the
extremal case is the existence of a well defined near-horizon geometry. Recently the general structure of such near-horizon geometries has been understood~\cite{Kunduri:2007vf}. We used this in our proof to determine the structure of the orbit space near the horizon. In fact, in 4d it had been shown
that such vacuum near-horizon geometries are
unique~\cite{Haj, LP, Kunduri:2008rs} (and specified by the angular
momentum). Therefore our Theorem 1 may be thought of as a generalisation of this result (although note that the near-horizon geometry uniqueness proof does not assume asymptotic flatness).

In $D=5$ the situation is more complicated for a number of reasons.
Firstly, as in the non-extremal case the interval structure is
non-trivial, and given an interval structure currently there is no
general way to know whether a corresponding regular black hole
solution exists (other than constructing it explicitly!). This is a
remaining crucial issue that must be settled in the non-extremal
case in order to achieve results as strong as in 4d (at least with
the assumed symmetries). Secondly, in the extremal case a new complication
arises as the near-horizon geometries of such black holes are not
unique. These have been classified (for non-toroidal horizons) and found to fall into three
classes: two types of $\Sp^3$ horizon geometry, say type A and type B, and one type of $\Sp^2 \times \Sp^1$ geometry~\cite{Kunduri:2008rs}.
Currently, it is not known what conditions are required in order for
a near-horizon geometry to arise as a near-horizon limit of a
(asymptotically flat) black hole solution. These kinds of problems
concern existence of a regular black hole solution given a set of data (the
near-horizon data, asymptotic flatness, the interval structure and
the angular momenta). New ideas will be required in order to answer
such questions. It is worth emphasising that, for these reasons, Theorem 2 is not a generalisation of the explicit classification of possible near-horizon geometries. It tells us that the near-horizon geometry of an (asymptotically flat) extremal black hole is uniquely determined by its angular momentum and interval structure -- this statement clearly has less content than the explicit classification of all possible near-horizon geometries.

In fact there are only two known explicit examples of 5d
asymptotically flat, stationary, extremal black holes: the extremal
Myers-Perry black hole~\cite{MP} and the extremal black
ring~\cite{Pomeransky:2006bd} both two parameter families uniquely
specified by their (non-vanishing) angular momenta $J_1,J_2$. The
near-horizon geometry of the extremal Myers-Perry is isometric to
the type A $\Sp^3$ horizon geometry discussed above. The
near-horizon geometry of the black ring corresponds to a special
case of the $\Sp^2 \times \Sp^1$ class. The type B $\Sp^3$ horizon
geometry and the most general ring topology case are not known to
arise as near-horizon limits of asymptotically flat black holes. An
interesting question which our work does not answer is whether there
are corresponding asymptotically flat black holes of this kind
(there are asymptotically KK ones). An argument against the
spherical topology class has been given in~\cite{Kunduri:2008rs}. As
for the ring topology case one can give a physical argument against
the existence of asymptotically flat black ring solutions with more
general near-horizon geometry. This is because in the classification
of near-horizon geometries it was shown that the most general ring
topology horizon geometry corresponds (essentially) to the boosted
extremal Kerr-string. This object is tensionless for a particular
value of the boost, and its near-horizon geometry is in fact
isometric to that of the known extremal black ring. Therefore, if
there are more general solutions with $\Sp^2 \times \Sp^1$ horizon
geometries, they would correspond to the
near-horizon geometries of black strings with tension. This seems
unlikely as heuristically one can imagine constructing a black ring
from a black string by bending a string into a loop and this can
only be done (without inputting external forces) for a tensionless
string~\cite{Figueras:2008qh}.

We have only considered asymptotically flat black holes. However, in
5d vacuum spacetime may also be asymptotically Kaluza-Klein. The
most general such case is to have the KK $\Sp^1$ fibered over 4d
Minkowski space (\ie\ the KK monopole). In the non-extremal case such a uniqueness theorem has been proved in the case where the $\Sp^1$ at infinity is not fibered (\ie\ the asymptotics are the direct product of 4d Minkowski and $\Sp^1$)~\cite{Hollands:2008fm}.  It should be straightforward
to extend our results to these cases too, with the only change
involving the boundary conditions at spatial infinity. There are in
fact three known extremal KK vacuum black holes.\footnote{One also expects an extremal black ring which asymptotes to the KK monopole to exist. In the non-extremal a special case has been constructed~\cite{Camps:2008hb}.} There are two
$\Sp^3$ topology KK black hole which asymptote to the KK monopole,
one termed the slowly rotating solution and the other the fast
rotating one. The other solution is the boosted extremal Kerr string
which is asymptotically the direct product of 4d Minkowski and
$\Sp^1$. Interestingly, in contrast to the asymptotically flat case, all the possible vacuum near-horizon geometries classified in~\cite{Kunduri:2008rs} actually correspond to those of known extremal KK vacuum black holes.\footnote{ The two
parameter family type A $\Sp^3$ case is also isometric to the slow extremal limit of the KK black hole, the three parameter type B $\Sp^3$ case includes the fast extremal limit of the KK black hole and the three parameter $\Sp^2 \times \Sp^1$ case includes the boosted extremal Kerr string.} 

The uniqueness theorem that we have proved could also be extended
without difficulty to stationary black hole spacetimes with
disconnected horizons, with at least one degenerate component. In
4d, regular spacetimes of this kind are not expected to exist
(indeed even regular stationary non-extremal multi black holes are
not expected). In 5d though, this type of spacetime should exist,
although we are not aware of any explicit example. For instance,
consider a five dimensional solution consisting of two doubly
spinning black holes, one of which is a black ring (\eg,  a doubly
spinning version of the black saturn~\cite{Elvang:2007rd} or the various  known multi-rings~\cite{biring}). One expects such configurations to admit
 extremal limits in which either one or both the horizons become
degenerate. This is because vacuum black rings (either non-extremal or extremal) can be made
arbitrarily thin and therefore the gravitational field at the centre
of the ring can be made arbitrarily weak. Therefore, one could
consider putting a sufficiently small (in comparison with the radius
of the ring) doubly spinning black hole at the centre and the
gravitational attraction between the two objects would be
arbitrarily small. Furthermore, it should be possible to tune the
angular momenta of the central object until it becomes
extremal, while equilibrium is maintained.  Clearly this argument
can also be applied to spacetimes containing multiple black rings. For such spacetimes, the orbit space would contain
the asymptotic end corresponding to spatial infinity together with
as many asymptotic ends as degenerate horizons (each corresponding
to a near-horizon geometry). Therefore, it would also contain
multiple disconnected boundaries, each of which in turn would be
divided into intervals. Once the multiple asymptotic ends and
boundaries are accounted for, the proof of the theorem should go
through unchanged.

It would also be interesting to extend these theorems to theories
containing charged black holes. This is realistic in
four-dimensional Einstein-Maxwell or five-dimensional minimal
supergravity as they are known to have non-linear sigma model descriptions
\cite{Mazur:1982db,Bouchareb:2007ax,Gal'tsov:2008nz}. Therefore, the
Mazur argument can be applied and an analogous uniqueness theorem
follows once the boundary conditions for the gauge fields are
properly taken into account. Indeed~\cite{Mazur:1982db} proved the uniqueness of the non-extremal Kerr-Newman solution, and recently
an analogous result has been shown in the context of 5d
minimal supergravity~\cite{Tomizawa:2009ua} (see also~\cite{Hollands:2007qf} for a result in 5d pure Einstein-Maxwell system). In the extremal case one would again require an understanding of the form of the near-horizon geometry. In 4d these have been
classified and found to be given by the near-horizon limit of the extremal
Kerr-Newman black hole~\cite{LP,Kunduri:2008tk} and thus a
uniqueness theorem for this case will be straightforward. In 5d such near-horizon geometries have not been classified (they have for the special case of supersymmetric black holes~\cite{Reall:2002bh}) and in
fact this appears to be a difficult problem in itself~\cite{KL}. However, we do have a general understanding of the structure of such near-horizon geometries~\cite{Kunduri:2007vf} which should be sufficient to prove a uniqueness theorem analogous to our Theorem 2 (as this does not require one to specify the explicit near-horizon geometry).

%%%%%%%%%%%%%%%%%%%%%%%%%%%%%%%%%%%%%%%%%%%%%%%%%%
%____________________________________________
\subsection*{Acknowledgements}
\label{sec:acks}
%____________________________________________

We would like to thank Gary Horowitz for helpful discussions.  PF and JL are supported by an STFC Rolling Grant.

%__________________________________________________
\appendix

\section{Staticity theorem}
\label{appendix:static}
In this appendix we provide the proofs of  the Staticity Theorem and Remark \ref{remark:multistatic} following it, which can be found in section \ref{sec:Preliminaries}. 

\begin{proof}[Proof of staticity theorem:]
 First note the following identity on $\mathcal{B}$:
\be
D^a \left( \frac{\rho \lambda^{ij} Y_i D_a Y_j}{\det \lambda} \right)= \frac{\rho}{\det \lambda } \lambda^{ij} D^a Y_i D_a Y_j- Y_i D^a \left(\frac{\rho}{\det \lambda} \, \lambda^{ij} \, D_a Y_j \; 
\right) \; .
\ee
Using the Einstein equations (\ref{lambdaYeqs}) and integrating over $\mathcal{B}$ gives
\be
\label{staticid}
\int_{\mathcal{B}} \frac{\rho}{\det \lambda } \lambda^{ij} D^a Y_i D_a Y_j = \int_{\partial \mathcal{B}} \frac{\rho}{\det \lambda} \lambda^{ij} Y_i \star_2 dY_j
\ee
where $\partial \mathcal{B}$ denotes the boundary of $\mathcal{B}$ (including any asymptotic regions). As discussed in \sec{sec:orbit}, $\partial \mathcal{B}$ consists of $\rho=0$ (which corresponds to the horizon and the axes) and asymptotic infinity $\rho\to \infty$. In order to evaluate the boundary integral on the RHS of (\ref{staticid}) it is helpful to note that, using (\ref{domegaD}), it can be written as $-\int_{\partial \mathcal{B}} Y_id \omega^i$. For asymptotically flat spacetimes $\omega^i \to 0$ and $Y_i$ are bounded as $\rho \to \infty$ and hence the contribution to the boundary integral from this region must vanish. Further, as argued in section (\ref{sec:Preliminaries}), on the horizon $d\omega^i=0$ and the $Y_i$ are well behaved. Therefore the contribution to the boundary integral from the horizon must also vanish (this argument is valid in both the non-extremal and extremal cases). This leaves to evaluate $-\int_{\mathcal{A}} Y_i d\omega^i$. For a connected horizon $\mathcal{A}= \mathcal{A}^+ \cup \mathcal{A}^-$, where $\mathcal{A}^{\pm}$ are the two axes which extend to infinity, and thus
\be
-\int_{\mathcal{A}} Y_i d\omega^i = -Y^+_i \int_{\mathcal{A}^+} d\omega^i-Y^-_i \int_{\mathcal{A}^-} d\omega^i= \omega_H^i(Y^+_i-Y^-_i) = \frac{8}{(2\pi)^{D-4}} \omega_H^i J_i
\label{eq:intstat}
\ee
where $\omega^i_H$ is the angular velocity of the horizon, and $Y^{\pm}_i=\pm 4\, J_i / (2\pi)^{D-4}$ (see section \ref{sec:axes}). Therefore, the RHS of (\ref{staticid}) is equal to $8\omega_H^i J_i/(2\pi)^{D-4}$. Since the integrand on the LHS of (\ref{staticid}) is positive definite, we deduce that $\omega_H^iJ_i=0$ if and only if $D_a Y_i=0$ everywhere in $\mathcal{B}$. But, $Y_i=\textrm{const}$ is equivalent to $\omega^i=\textrm{const}$ which is equivalent to $\xi=\partial / \partial t$ being hypersurface orthogonal (since the asymptotic condition $\omega^i \to 0$ fixes $\omega^i \equiv 0$). This  completes the proof.
\end{proof}

\begin{proof}[Proof of Remark \ref{remark:multistatic}]  To prove the last statement in this remark we proceed as in the proof above, we argue that the contribution to the RHS of \eqref{staticid} coming from the boundary at infinity and the $N$  components of the horizon vanishes.  Next we consider   the contribution from the axes set. However,  now   one has to bear in mind that since for a non-connected horizon $\mathcal{A}$  has more than two disconnected components, at least one of them is necessarily not connected to asymptotic infinity. Therefore, one cannot fix the value of $Y_i$ in \eqref{eq:intstat} to be proportional to $J_i$ on these parts of the axes set. However,  the integral $-\int_{\mathcal{A}} Y_i d\omega^i$ can be evaluated in general in this case too. Suppose that there are $N$ disconnected components of the horizon. This implies that $\mathcal{A}$ is a disjoint union of $N+1$ disconnected sets $\mathcal{A}^I$, with $I=0,1 \dots N$, and define the constants $Y^I_i \equiv Y_i|_{\mathcal{A}^I}$. Therefore
\be
-\int_{\mathcal{A}} Y_i d\omega^i= -\sum_{I=0}^N Y^I_i \int_{\mathcal{A}^I} d\omega^i =  \sum_{I=0}^N Y^I_i (\omega_{I+1}^i-\omega_{I}^i)= \sum_{I=1}^N (Y^{I-1}_i-Y^I_i)\omega_I^i
\ee
where $\omega_I^i$ for $I=1, \dots, N$ are the angular velocities of the $I^{\textrm{th}}$ component of the horizon and $\omega_{0}^i=\omega_{N+1}^i=0$. Now in general\footnote{This expression is a coordinate-independent generalisation of the relation derived in \cite{Hollands:2007aj} valid for both non-degenerate and degenerate horizons.} 
\begin{equation}
\begin{aligned}
Y^{I-1}_i-Y^I_i =&~~\int_{\mathcal{H}_I/U(1)^{D-3}} \Omega_i\\
=&~~\frac{1}{(2\pi)^{D-3}}\int_{\mathcal{H}_I} \star dm_i = \frac{16\pi}{(2\pi)^{D-3}}\,J^I_i\,,
\end{aligned}
\end{equation}
 where $J^I_i$ are the Komar angular momenta of the $I^{\textrm{th}}$ component of the horizon. In going from the first to the second line we have used the identity $(\star dm_i)|_{\mathcal{H}}= (\Omega_i \wedge d\phi^1\wedge \cdots \wedge d\phi^{D-3})|_{\mathcal{H}}$ and the fact $L_{m_i} \Omega_j=0$ (\ie\ $\Omega_i$ are 1-forms on the orbit space) to convert the integral over the $I^{\textrm{th}}$ horizon orbit space to an (spacetime) integral over the $I^{\textrm{th}}$ horizon. We have thus derived the general identity
\be
\int_{\mathcal{B}} \frac{\rho}{\det \lambda } \lambda^{ij} D^a Y_i D_a Y_j = \frac{8}{(2\pi)^{D-4}} \sum_{I=1}^N \omega_I^i J^I_i \; .
\ee
Therefore, the spacetime is static if and only if $\sum_{I=1}^N \omega_I^i J^I_i=0$. This completes the proof. 
\end{proof}

\section{Weyl solutions in spherical polar coordinates}
%__________________________________________________
\label{subsec:sphpolarcoords} Assume the event horizon has a single
connected component with zero surface gravity. As shown in the text we can use spherical polar coordinates
\begin{equation}
 \rho=r\,\sin\theta\,,\qquad z=r\,\cos\theta\,,
\end{equation}
where $0 \leq \theta \leq \pi$ (since $\rho \geq 0$), so that the
horizon now is located at the origin $r \to 0$. These coordinates are
valid everywhere for $r>0$, even away from the near-horizon limit.
The results in \S\ref{sec:orbit} show that they coincide near the
horizon (strictly in the near-horizon limit) with the coordinates
originating from Gaussian null coordinates.  In this Appendix we
have written Einstein equations \eqref{Gequation} and
\eqref{Riccibase} in these spherical polar coordinates:
\begin{subequations}
\begin{align}
&\left[\partial_r(r^2\,\partial_r)
    +\frac{1}{\sin\theta}\,\partial_\theta\left(\sin\theta\,\partial_\theta\right)
    \right]\lambda_{ij}=\lambda^{mn}
    \left[r^2\,(\partial_r\lambda_{im})(\partial_r\lambda_{jn})+
    (\partial_\theta\lambda_{im})(\partial_\theta\lambda_{jn})\right]\nonumber\\
&\hspace{6.5cm}-\frac{1}{\det\lambda}
    \left[r^2\,(\partial_rY_{i})(\partial_rY_{j})+
    (\partial_\theta Y_{i})(\partial_\theta Y_{j})\right]\,,\\
&\left[\partial_r(r^2\,\partial_r)
    +\frac{1}{\sin\theta}\,\partial_\theta\left(\sin\theta\,\partial_\theta\right)
    \right]Y_{i}=\lambda^{mn}
    \left[r^2\,(\partial_r\lambda_{im})(\partial_rY_{n})+
    (\partial_\theta\lambda_{im})(\partial_\theta Y_{n})\right]\nonumber\\
&\hspace{6.4cm}+r^2\,(\partial_r\ln\det\lambda)(\partial_rY_i)
    +(\partial_\theta\ln\det\lambda)(\partial_\theta Y_i)\,,
\end{align}
\end{subequations}

\begin{eqnarray}
%\begin{align}
&\partial_r\nu=-\frac{1}{2}\,\partial_r\ln\det\lambda+\frac{1}{8}\,\bigg\{
r\,\sin^2\theta\bigg[(\partial_r\ln\det\lambda)^2
    -\frac{1}{r^2}\,(\partial_\theta\ln\det\lambda)^2\nonumber\\
&\hspace{6.3cm}+\textrm{Tr}\big((\lambda^{-1}\partial_r\lambda)^2\big)
    -\frac{1}{r^2}\,\textrm{Tr}\big((\lambda^{-1}\partial_\theta\lambda)^2\big)
    \nonumber\\
&\hspace{6.3cm}+\frac{2}{\det\lambda}\,\lambda^{ij}\Big(
    (\partial_rY_i)(\partial_rY_j)
    -\frac{1}{r^2}\,(\partial_\theta Y_i)(\partial_\theta Y_j)\Big)\bigg]\nonumber\\
&\hspace{4.8cm}+2\,\sin\theta\,\cos\theta\bigg[
(\partial_r\ln\det\lambda)(\partial_\theta\ln\det\lambda)
+\textrm{Tr}\big((\lambda^{-1}\partial_r\lambda)(\lambda^{-1}\partial_\theta\lambda)\big)
\nonumber\\
&\hspace{7.6cm}+\frac{2}{\det\lambda}\,\lambda^{ij}(\partial_r Y_i)(\partial_\theta Y_j)
\bigg]
\bigg\}\,,
%\end{align}
\end{eqnarray}
\begin{eqnarray}
%\begin{align}
&\partial_\theta\nu=-\frac{1}{2}\,\partial_\theta\ln\det\lambda-\frac{1}{8}\,\bigg\{
r^2\sin\theta\,\cos\theta\bigg[
(\partial_r\ln\det\lambda)^2-\frac{1}{r^2}\,(\partial_\theta\ln\det\lambda)^2\nonumber\\
&\hspace{7.2cm}+\textrm{Tr}\big((\lambda^{-1}\partial_r\lambda)^2\big)
    -\frac{1}{r^2}\,\textrm{Tr}\big((\lambda^{-1}\partial_\theta\lambda)^2\big)
\nonumber\\
&\hspace{7.2cm}+\frac{2}{\det\lambda}\,\lambda^{ij}\Big(
    (\partial_rY_i)(\partial_rY_j)
    -\frac{1}{r^2}\,(\partial_\theta Y_i)(\partial_\theta Y_j)\Big)\bigg]\nonumber\\
&\hspace{4.8cm}-2\,r\,\sin^2\theta\bigg[
(\partial_r\ln\det\lambda)(\partial_\theta\ln\det\lambda)
+\textrm{Tr}\big((\lambda^{-1}\partial_r\lambda)(\lambda^{-1}\partial_\theta\lambda)\big)
\nonumber\\
&\hspace{7.1cm}+\frac{2}{\det\lambda}\,\lambda^{ij}(\partial_r Y_i)(\partial_\theta Y_j)
\bigg]
\bigg\}
%\end{align}
\end{eqnarray}
Also note
\begin{equation}
\partial_r \omega_i = -\frac{\sin\theta}{\det \lambda} \lambda^{ij}
\partial_\theta Y_j, \qquad \partial_\theta \omega_i =
\frac{r^2\sin\theta}{\det \lambda} \lambda^{ij}\partial_r Y_j \; .
\end{equation}

%%%%%%%%%%%%%%%%%%%%%%%%%%%%%%%%%%%%%%%%%%%%

%\endgroup

\begin{thebibliography}{99}
{\small

\bibitem{Carter:1971zc}
  B.~Carter,
  ``Axisymmetric Black Hole Has Only Two Degrees of Freedom,''
  Phys.\ Rev.\ Lett.\  {\bf 26} (1971) 331.

\bibitem{Carter:1973lh}
  B.~Carter,
  ``Black Hole Equilibrium States: II General Theory of Stationary Black Hole States,''
  in \textit{Black Holes} (proc. 1972 Les Houches Summer School), ed. B. $\&$ C. DeWitt,
  125-210 (Gordon and Breach, New York, 1973).

%\cite{Robinson:1975bv}
\bibitem{Robinson:1975bv}
  D.~C.~Robinson,
  ``Uniqueness of the Kerr black hole,''
  Phys.\ Rev.\ Lett.\  {\bf 34} (1975) 905.
  %%CITATION = PRLTA,34,905;%%
%\cite{Mazur:1982db}

\bibitem{Mazur:1982db}
  P.~O.~Mazur,
  ``Proof Of Uniqueness Of The Kerr-Newman Black Hole Solution,''
  J.\ Phys.\ A  {\bf 15} (1982) 3173.
  %%CITATION = JPAGB,A15,3173;%%

\bibitem{Mazur:1983dc}
  P.~O.~Mazur,
  ``Black Hole Uniqueness From A Hidden Symmetry Of Einstein's Gravity,''
  Gen.\ Rel.\ Grav.\  {\bf 16} (1984) 211.


\bibitem{Bunting:1983bu}
  G.~Bunting,
  ``Proof of the uniqueness conjecture for black holes.''
  PhD Thesis, Department of Mathematics, University of New England, Armidale, N.S.W. 1983.

\bibitem{Carter:1985bu}
  B.~Carter,
  ``Bunting Identity and Mazur Identity for Non-linear Elliptic Systems Including the Black Hole Equilibrium Problem,''
  Commun.\ Math.\ Phys.\  {\bf 99} (1985) 563.

%\cite{Chrusciel:2008js}
\bibitem{Chrusciel:2008js}
  P.~T.~Chrusciel and J.~Lopes Costa,
  ``On uniqueness of stationary vacuum black holes,''
  arXiv:0806.0016 [gr-qc].
  %%CITATION = ARXIV:0806.0016;%%

%\cite{Hawking:1971vc}
\bibitem{Hawking:1971vc}
  S.~W.~Hawking,
  ``Black holes in general relativity,''
  Commun.\ Math.\ Phys.\  {\bf 25} (1972) 152.
  %%CITATION = CMPHA,25,152;%%

%\cite{Hawking:1973uf}
\bibitem{Hawking:1973uf}
  S.~W.~Hawking and G.~F.~R.~Ellis,
  ``The Large scale structure of space-time,''
%\href{http://www.slac.stanford.edu/spires/find/hep/www?irn=6991262}{SPIRES entry}
{\it  Cambridge University Press, Cambridge, 1973}

\bibitem{Chrusciel:1994tr}
  P.~T.~Chrusciel and R.~M.~Wald,
  ``On The Topology Of Stationary Black Holes,''
  Class.\ Quant.\ Grav.\  {\bf 11} (1994) L147
  [arXiv:gr-qc/9410004].


%\cite{Galloway:1999bp}
\bibitem{Galloway:1999bp}
  G.~J.~Galloway, K.~Schleich, D.~M.~Witt and E.~Woolgar,
  ``Topological Censorship and Higher Genus Black Holes,''
  Phys.\ Rev.\  D {\bf 60} (1999) 104039
  [arXiv:gr-qc/9902061].
  %%CITATION = PHRVA,D60,104039;%%


\bibitem{Sudarsky:1992ty}
  D.~Sudarsky and R.~M.~Wald,
  ``Extrema of mass, stationarity, and staticity, and solutions to the Einstein
  Yang-Mills equations,''
  Phys.\ Rev.\  D {\bf 46} (1992) 1453.


%\cite{Israel:1967wq}
\bibitem{Israel:1967wq}
  W.~Israel,
  ``Event Horizons In Static Vacuum Space-Times,''
  Phys.\ Rev.\  {\bf 164} (1967) 1776.
  %%CITATION = PHRVA,164,1776;%%

\bibitem{BM}
G.L.~Bunting, A.K.M.~Masood-ul-Alam, ``Nonexistence of multiple black holes in asymptotically Euclidean static vacuum space-times'', Gen. rel. Grav.{\bf 19} (1987), 147-154


\bibitem{Chrusciel:1996bj}
  P.~T.~Chrusciel,
  ``On rigidity of analytic black holes,''
  Commun.\ Math.\ Phys.\  {\bf 189} (1997) 1
  [arXiv:gr-qc/9610011].

\bibitem{Friedrich:1998wq}
  H.~Friedrich, I.~Racz and R.~M.~Wald,
  ``On the Rigidity Theorem for Spacetimes with a Stationary Event Horizon or a
  Compact Cauchy Horizon,''
  Commun.\ Math.\ Phys.\  {\bf 204} (1999) 691
  [arXiv:gr-qc/9811021].


%\cite{Ernst:1967wx}
\bibitem{Ernst:1967wx}
  F.~J.~Ernst,
  ``New formulation of the axially symmetric gravitational field problem,''
  Phys.\ Rev.\  {\bf 167} (1968) 1175.
  %%CITATION = PHRVA,167,1175;%%

\bibitem{Mazur:1983vi}
  P.~O.~Mazur,
  ``A Relationship Between The Electrovacuum Ernst Equations And Nonlinear
  Sigma Model,''
  Acta Phys.\ Polon.\  B {\bf 14} (1983) 219.


\bibitem{Chrusciel:2005pa}
  P.~T.~Chrusciel, H.~S.~Reall and P.~Tod,
  ``On non-existence of static vacuum black holes with degenerate  components
  of the event horizon,''
  Class.\ Quant.\ Grav.\  {\bf 23} (2006) 549
  [arXiv:gr-qc/0512041].

%\cite{Hollands:2008wn}
\bibitem{Hollands:2008wn}
  S.~Hollands and A.~Ishibashi,
  ``On the `Stationary Implies Axisymmetric' Theorem for Extremal Black Holes
  in Higher Dimensions,''
  arXiv:0809.2659 [gr-qc].
  %%CITATION = ARXIV:0809.2659;%%

%\cite{Kunduri:2007vf}
\bibitem{Kunduri:2007vf}
  H.~K.~Kunduri, J.~Lucietti and H.~S.~Reall,
  ``Near-horizon symmetries of extremal black holes,''
  Class.\ Quant.\ Grav.\  {\bf 24} (2007) 4169
  [arXiv:0705.4214 [hep-th]].
  %%CITATION = CQGRD,24,4169;%%


\bibitem{Haj}
P.~H{a}j{{i}}{{c}}ek, ``Three remarks on axisymmetric stationary horizons'', Commun.Math.\ Phys. {\bf 36} (1974), p.~305--320.

\bibitem{LP}
  J.~Lewandowski and T.~Pawlowski,
  ``Extremal Isolated Horizons: A Local Uniqueness Theorem,''
  Class.\ Quant.\ Grav.\  {\bf 20} (2003) 587
  [arXiv:gr-qc/0208032].

%\cite{Kunduri:2008rs}
\bibitem{Kunduri:2008rs}
  H.~K.~Kunduri and J.~Lucietti,
  ``A classification of near-horizon geometries of extremal vacuum black
  holes,''
  arXiv:0806.2051 [hep-th].
  %%CITATION = ARXIV:0806.2051;%%

\bibitem{MP}
  R.~C.~Myers and M.~J.~Perry,
  ``Black Holes In Higher Dimensional Space-Times,''
  Annals Phys.\  {\bf 172}, 304 (1986).

%\cite{Emparan:2001wk}
\bibitem{Emparan:2001wk}
  R.~Emparan and H.~S.~Reall,
  ``Generalized Weyl solutions,''
  Phys.\ Rev.\  D {\bf 65} (2002) 084025
  [arXiv:hep-th/0110258].
  %%CITATION = PHRVA,D65,084025;%%

%\cite{Emparan:2001wn}
\bibitem{Emparan:2001wn}
  R.~Emparan and H.~S.~Reall,
  ``A rotating black ring in five dimensions,''
  Phys.\ Rev.\ Lett.\  {\bf 88} (2002) 101101
  [arXiv:hep-th/0110260].
  %%CITATION = PRLTA,88,101101;%%

%\cite{Pomeransky:2006bd}
\bibitem{Pomeransky:2006bd}
  A.~A.~Pomeransky and R.~A.~Sen'kov,
  ``Black ring with two angular momenta,''
  arXiv:hep-th/0612005.
  %%CITATION = HEP-TH/0612005;%%

\bibitem{Galloway:2005mf}
  G.~J.~Galloway and R.~Schoen,
  ``A generalization of Hawking's black hole topology theorem to higher
  dimensions,''
  Commun.\ Math.\ Phys.\  {\bf 266} (2006) 571
  [arXiv:gr-qc/0509107].

\bibitem{Galloway:2006ws}
  G.~J.~Galloway,
  ``Rigidity of outer horizons and the topology of black holes,''
  arXiv:gr-qc/0608118.


\bibitem{Gibbons:2002bh}
  G.~W.~Gibbons, D.~Ida and T.~Shiromizu,
  ``Uniqueness and non-uniqueness of static vacuum black holes in higher
  dimensions,''
  Prog.\ Theor.\ Phys.\ Suppl.\  {\bf 148} (2003) 284
  [arXiv:gr-qc/0203004].

\bibitem{Rogatko:2005ys}
  M.~Rogatko,
  ``Staticity theorem for higher dimensional generalized Einstein-Maxwell
  system,''
  Phys.\ Rev.\  D {\bf 71} (2005) 024031
  [arXiv:hep-th/0501216].


%\cite{Hollands:2006rj}
\bibitem{Hollands:2006rj}
  S.~Hollands, A.~Ishibashi and R.~M.~Wald,
  ``A Higher Dimensional Stationary Rotating Black Hole Must be Axisymmetric,''
  Commun.\ Math.\ Phys.\  {\bf 271} (2007) 699
  [arXiv:gr-qc/0605106].
  %%CITATION = CMPHA,271,699;%%

%\cite{Moncrief:2008mr}
\bibitem{Moncrief:2008mr}
  V.~Moncrief and J.~Isenberg,
  ``Symmetries of Higher Dimensional Black Holes,''
  Class.\ Quant.\ Grav.\  {\bf 25} (2008) 195015
  [arXiv:0805.1451 [gr-qc]].
  %%CITATION = CQGRD,25,195015;%%


%\cite{Harmark:2004rm}
\bibitem{Harmark:2004rm}
  T.~Harmark,
  ``Stationary and axisymmetric solutions of higher-dimensional general
  relativity,''
  Phys.\ Rev.\  D {\bf 70} (2004) 124002
  [arXiv:hep-th/0408141].
  %%CITATION = PHRVA,D70,124002;%%


\bibitem{Chrusciel:2008rh}
  P.~T.~Chrusciel,
  ``On higher dimensional black holes with abelian isometry group,''
  arXiv:0812.3424 [gr-qc].

%\cite{Hollands:2007aj}
\bibitem{Hollands:2007aj}
  S.~Hollands and S.~Yazadjiev,
  ``Uniqueness theorem for 5-dimensional black holes with two axial Killing
  fields,''
  Commun.\ Math.\ Phys.\  {\bf 283} (2008) 749
  [arXiv:0707.2775 [gr-qc]].
  %%CITATION = CMPHA,283,749;%%

\bibitem{Hollands:2008fm}
  S.~Hollands and S.~Yazadjiev,
  ``A uniqueness theorem for stationary Kaluza-Klein black holes,''
  arXiv:0812.3036 [gr-qc].

\bibitem{Gowdy}
  R.~H.~Gowdy,
  ``Vacuum space-times with two parameter spacelike isometry groups and compact
  invariant hypersurfaces: Topologies and boundary conditions,''
  Annals Phys.\  {\bf 83} (1974) 203.

\bibitem{Chrusciel2rot}
  P.~Chrusciel,
  ``On space-times with $U(1)\times U(1)$ symmetric compact Cauchy surfaces,''
  Annals Phys.\  {\bf 202} (1990) 100.


%\cite{Evslin:2008gx}
\bibitem{lenses}
  J.~Evslin,
  ``Geometric Engineering 5d Black Holes with Rod Diagrams,''
  JHEP {\bf 0809} (2008) 004
  [arXiv:0806.3389 [hep-th]].
  %%CITATION = JHEPA,0809,004;%%

  %\cite{Chen:2008fa}
%\bibitem{Chen:2008fa}
  Y.~Chen and E.~Teo,
  ``A rotating black lens solution in five dimensions,''
  Phys.\ Rev.\  D {\bf 78} (2008) 064062
  [arXiv:0808.0587 [gr-qc]].
  %%CITATION = PHRVA,D78,064062;%%

\bibitem{Weinstein1}
G.~Weinstein, ``On rotating black holes in equilibrium in General Relativity'',  Commun.Pure Appl.Math. {\bf 43}, 903 (1990)

\bibitem{Weinstein2}
G.~Weinstein, ``On the Dirichlet problem for harmonic maps with prescribed singularities'', Duke Math.J. {\bf 77}, 135 (1995) No.1 135-165

\bibitem{Chrusciel:2007ak}
  P.~T.~Chrusciel, Y.~Li and G.~Weinstein,
  ``Mass and angular-momentum inequalities for axi-symmetric initial data sets.
  II. Angular-momentum,''
  Annals Phys.\  {\bf 323} (2008) 2591
  [arXiv:0712.4064 [gr-qc]].

\bibitem{Zaslavsky:1997uu}
  O.~B.~Zaslavsky,
  ``Horizon/Matter Systems Near the Extreme State,''
  Class.\ Quant.\ Grav.\  {\bf 15} (1998) 3251
  [arXiv:gr-qc/9712007].


%\cite{Bardeen:1999px}
\bibitem{Bardeen:1999px}
  J.~M.~Bardeen and G.~T.~Horowitz,
  ``The extreme Kerr throat geometry: A vacuum analog of AdS(2) x S(2),''
  Phys.\ Rev.\  D {\bf 60} (1999) 104030
  [arXiv:hep-th/9905099].
  %%CITATION = PHRVA,D60,104030;%%


%\cite{Figueras:2008qh}
\bibitem{Figueras:2008qh}
  P.~Figueras, H.~K.~Kunduri, J.~Lucietti and M.~Rangamani,
  ``Extremal vacuum black holes in higher dimensions,''
  Phys.\ Rev.\  D {\bf 78} (2008) 044042
  [arXiv:0803.2998 [hep-th]].
  %%CITATION = PHRVA,D78,044042;%%

\bibitem{Camps:2008hb}
  J.~Camps, R.~Emparan, P.~Figueras, S.~Giusto and A.~Saxena,
  ``Black Rings in Taub-NUT and D0-D6 interactions,''
  JHEP {\bf 0902} (2009) 021
  [arXiv:0811.2088 [hep-th]].



%\cite{Elvang:2007rd}
\bibitem{Elvang:2007rd}
  H.~Elvang and P.~Figueras,
  ``Black Saturn,''
  JHEP {\bf 0705} (2007) 050
  [arXiv:hep-th/0701035].
  %%CITATION = JHEPA,0705,050;%%

\bibitem{biring}
%\cite{Iguchi:2007is}
%\bibitem{Iguchi:2007is}
  H.~Iguchi and T.~Mishima,
  ``Black di-ring and infinite nonuniqueness,''
  Phys.\ Rev.\  D {\bf 75} (2007) 064018
  [Erratum-ibid.\  D {\bf 78} (2008) 069903]
  [arXiv:hep-th/0701043].
  %%CITATION = PHRVA,D75,064018;%%

%\cite{Evslin:2007fv}
%\bibitem{Evslin:2007fv}
  J.~Evslin and C.~Krishnan,
  %``The Black Di-Ring: An Inverse Scattering Construction,''
  Class.\ Quant.\ Grav.\  {\bf 26} (2009) 125018
  [arXiv:0706.1231 [hep-th]].
  %%CITATION = CQGRD,26,125018;%%

  K.~Izumi,
  ``Orthogonal black di-ring solution,''
  Prog.\ Theor.\ Phys.\  {\bf 119} (2008) 757
  [arXiv:0712.0902 [hep-th]].
  %%CITATION = PTPKA,119,757;%%

%\cite{Elvang:2007hs}
%\bibitem{biring}
  H.~Elvang and M.~J.~Rodriguez,
  ``Bicycling Black Rings,''
  JHEP {\bf 0804} (2008) 045
  [arXiv:0712.2425 [hep-th]].
  %%CITATION = JHEPA,0804,045;%%

%\cite{Bouchareb:2007ax}
\bibitem{Bouchareb:2007ax}
  A.~Bouchareb, G.~Clement, C.~M.~Chen, D.~V.~Gal'tsov, N.~G.~Scherbluk and T.~Wolf,
  ``$G_2$ generating technique for minimal D=5 supergravity and black rings,''
  Phys.\ Rev.\  D {\bf 76} (2007) 104032
  [Erratum-ibid.\  D {\bf 78} (2008) 029901]
  [arXiv:0708.2361 [hep-th]].
  %%CITATION = PHRVA,D76,104032;%%

%\cite{Gal'tsov:2008nz}
\bibitem{Gal'tsov:2008nz}
  D.~V.~Gal'tsov and N.~G.~Scherbluk,
  ``Generating technique for $U(1)^3 5D$ supergravity,''
  Phys.\ Rev.\  D {\bf 78} (2008) 064033
  [arXiv:0805.3924 [hep-th]].
  %%CITATION = PHRVA,D78,064033;%%

%\cite{Tomizawa:2009ua}
\bibitem{Tomizawa:2009ua}
  S.~Tomizawa, Y.~Yasui and A.~Ishibashi,
  ``A uniqueness theorem for charged rotating black holes in five-dimensional
  minimal supergravity,''
  arXiv:0901.4724 [hep-th].
  %%CITATION = ARXIV:0901.4724;%%

%\cite{Hollands:2007qf}
\bibitem{Hollands:2007qf}
  S.~Hollands and S.~Yazadjiev,
  ``A Uniqueness theorem for 5-dimensional Einstein-Maxwell black holes,''
  Class.\ Quant.\ Grav.\  {\bf 25} (2008) 095010
  [arXiv:0711.1722 [gr-qc]].
  %%CITATION = CQGRD,25,095010;%%

\bibitem{Kunduri:2008tk}
  H.~K.~Kunduri and J.~Lucietti,
  ``Uniqueness of near-horizon geometries of rotating extremal AdS(4) black
  holes,''
  Class.\ Quant.\ Grav.\  {\bf 26} (2009) 055019
  [arXiv:0812.1576 [hep-th]].

\bibitem{Reall:2002bh}
  H.~S.~Reall,
  ``Higher dimensional black holes and supersymmetry,''
  Phys.\ Rev.\  D {\bf 68} (2003) 024024
  [Erratum-ibid.\  D {\bf 70} (2004) 089902]
  [arXiv:hep-th/0211290].


\bibitem{KL}
H.~K.~Kunduri and J.~Lucietti,
  ``Static near-horizon geometries in five dimensions,''
  arXiv:0907.0410 [hep-th].

%\cite{Amsel:2009et}
\bibitem{Amsel:2009et}
  A.~J.~Amsel, G.~T.~Horowitz, D.~Marolf and M.~M.~Roberts,
  ``Uniqueness of Extremal Kerr and Kerr-Newman Black Holes,''
  arXiv:0906.2367 [gr-qc].
  %%CITATION = ARXIV:0906.2367;%%


\bibitem{Meinel}
R. Meinel, M. Ansorg, A. Kleinwachter, G. Neugebauer and D.
Petroff, 
``Relativistic Figures of
Equilibrium'', Cambridge University Press 2008 (sec 2.4)


}

\end{thebibliography}
\end{document}